\documentclass[prl,twocolumn,showpacs,superscriptaddress]{revtex4-2}
\usepackage{graphicx}
\usepackage{mathrsfs}
\usepackage{dcolumn}
\usepackage{bm}
\usepackage{amsmath}
\usepackage{amsfonts}
\usepackage{color}
\usepackage{makecell}
\usepackage{booktabs}
\usepackage{amssymb}
\usepackage{hyperref}
\hypersetup{hypertex=true,
	colorlinks=true,
	linkcolor=blue,
	anchorcolor=blue,
	citecolor=blue}

\begin{document}

\title{Electronic and topological characters of the ideal magnetic topological materials EuAuX with X = P, As, Sb, and Bi}
\author{Shengwei Chi}
\affiliation{Wuhan National High Magnetic Field Center $\&$ School of Physics, Huazhong University of Science and Technology, Wuhan 430074, China}
\author{Gang Xu}
\email[e-mail address: ]{gangxu@hust.edu.cn}
\affiliation{Wuhan National High Magnetic Field Center $\&$ School of Physics, Huazhong University of Science and Technology, Wuhan 430074, China}
\affiliation{Institute for Quantum Science and Engineering, Huazhong University of Science and Technology, Wuhan, 430074, China}
\affiliation{Wuhan Institute of Quantum Technology, Wuhan, 430074, China}

\begin{abstract}

Ideal magnetic topological materials have great significance in both fundamental physics and technical applications, due to their abundant exotic quantum properties and facilitation of control. 
Using first-principles calculations, we find several ideal magnetic topological materials in EuTX (T = Cu, Ag, and Au; X= P, As, Sb, and Bi) family. 
Particularly, EuAuP is the ferromagnetic Weyl semimetal, and EuAuX (X=As, Sb, and Bi) in their ground state with in-plane moments are the antiferromagnetic semimetals hosting the topological gap near the Fermi level. 
By tuning the magnetic moments to $z$-axis, EuAuX (X=As, Sb, and Bi) could further evolve into triple degenerate nodal points (TDNPs) semimetal states. 
The main characteristics of antiferromagnetic TDNP semimetal, including the Fermi arcs, and tangent Fermi surfaces with opposite spin winding numbers, are also studied. 
Our work provides a promising platform to modulate the magnetism, topological electronic structures and emergent quantum states.

\end{abstract}

\maketitle

\renewcommand\thesection{\Roman{section}}

\section{INTRODUCTION}

Topological matters have attracted extensive research attention during almost two decades\cite{hasan2010colloquium,qi2011topological,chiu2016classification,armitage2018weyl}, including quantum spin Hall (QSH) effect \cite{kane2005quantum,bernevig2006quantum} , three-dimensional topological insulator \cite{teo2008surface,zhang2009topological,yan2010theoretical} and topological semimetal (TSM) \cite{wan2011topological,xu2011chern,burkov2011weyl,zyuzin2012weyl,young2012dirac,wang2012dirac,wang2013three,yang2014classification,li2022topological}. 
In particular, TSM is classified by the degree of degeneracy of the band crossing points near the Fermi level. 
For instance, Weyl semimetals are characterized by twofold degenerate points near the Fermi level, and their low-energy excitations satisfy the Weyl equation. 
Analogously, Dirac semimetals are characterized by two doubly degenerate bands that cross to form fourfold degenerate points \cite{burkov2016topological}.
Meanwhile, some unconventional quasiparticle excitations beyond relativistic quantum field theory emerge in condensed matter, such as linear and quadratic three-, six-, and eightfold band crossings \cite{weng2016topological,bradlyn2016beyond,tang2017multiple,zhu2016triple,wang2017prediction,mondal2019broken}.
These topological matters exhibit exotic properties such as backscattering suppression \cite{buttiker1988absence}, spin-momentum locking \cite{hsieh2009tunable}, Fermi arcs \cite{weng2015weyl,lv2015experimental}, and negative magnetoresistance \cite{li2016negative}, which is prospective for spin manipulations and spintronic applications \cite{mellnik2014spin}.

Recently, research on topological matters has been expanded to magnetic materials \cite{tang2016dirac,hua2018dirac,zhang2019topological,zou2019study,jin2021multiple,malick2022electronic}. 
Due to the broken  time reversal symmetry, the system can exhibit new types of topological properties, such as quantum anomalous Hall (QAH) effect \cite{chang2023colloquium} and axion electrodynamics \cite{xu2019higher}. 
For example, MnBi$_2$Te$_4$ \cite{zhang2019topological,deng2020quantum} is lately discovered to be intrinsic magnetic topological insulators to realize the QAH effect and the Quantized Magnetoelectric effect.
Besides, the topological semimetal states including Dirac and Weyl semimetals can also appear in the magnetic system \cite{liu2019magnetic}.
For example, Dirac points can emerge in antiferromagnetic semimetal EuCd$_2$As$_2$ \cite{hua2018dirac}, which is protected by $\mathcal{P}$ and the nonsymmorphic time reversal symmetry $\mathcal{T}'$ in the magnetic space groups (MSGs).
Particularly, by tuning the magnetic ordering, EuCd$_2$As$_2$ can be transformed from Dirac semimetal into Weyl semimetal \cite{ma2019spin} or magnetic topological insulator.
While magnetic topological materials have great potential in physical property regulation and spintronic device development compared with nonmagnetic materials \cite{vsmejkal2018topological,he2022topological,bernevig2022progress}, ideal magnetic topological materials with obvious topological features near the Fermi level are still rare. 
Therefore, using $ab$ $initio$ calculation to find more magnetic topological materials is necessary.

In this paper, we use first-principles calculations based on HSE functional to study electronic structures and topological properties in the magnetic materials EuTX with T = Cu, Ag, and Au; X= P, As, Sb, and Bi systematically.  
Our results show that EuAgP, EuCuAs and EuAgAs are trivial insulators, EuCuP, EuCuSb and EuAgSb are trivial metals, while other materials are topological semimetals.
Particularly, EuAuP is ferromagnetic (FM) Weyl semimetal, and EuAuX (X = As, Sb, and Bi) are antiferromagnetic (AFM) semimetals with topological nontrivial gap near the Fermi level. 
The magnetic, electronic and topological properties of EuAuBi are studied in detail. 
Our calculations demonstrate that, in consistent with the experiments \cite{takahashi2023superconductivity}, the ground state of EuAuBi is the interlayer AFM with in-plane magnetic moments. 
Its corresponding electronic character is the AFM semimetal with a continuous topological gap near the Fermi level, which gives rise to a pair of gapped topological surface states existing just below the Fermi level. 
Since the energy of AFM EuAuBi with $z$-axis moments is only 0.2 meV/f.u. higher than the ground state, one can easily tune it into the AFM-$z$ configuration by an external magnetic field, and make EuAuBi evolve into the triple degenerate nodal points (TDNPs) semimetal state. 
The main characteristics of AFM TDNP semimetal, including the Fermi arcs and spin textures on the tangent Fermi surfaces, are also evidenced.

\begin{figure}[htbp]
	\centering
	\includegraphics[width=0.46\textwidth]{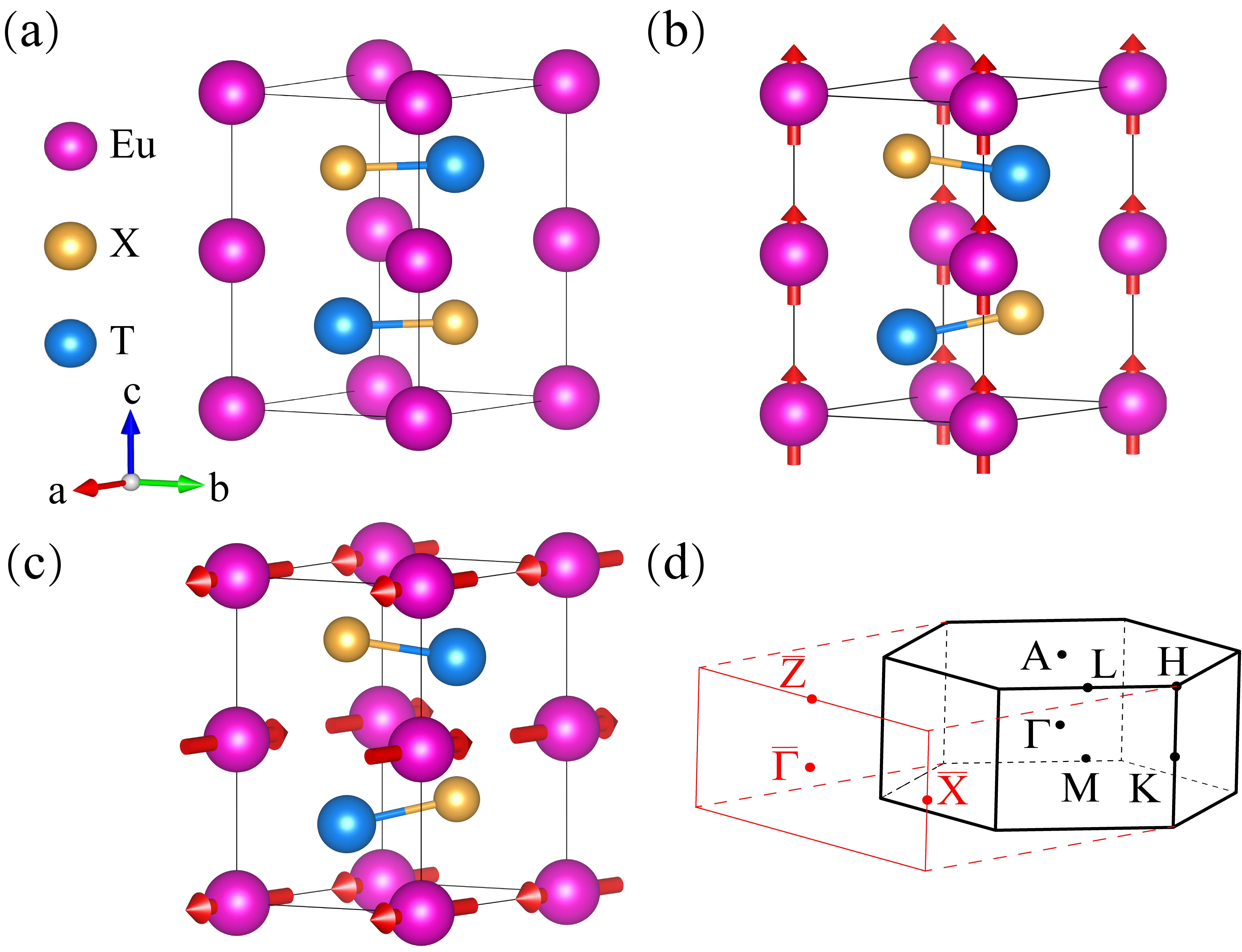}
	\caption{ (a) Crystal structure of EuTX in P6$_{3}/mmc$ space group. (b) Crystal structure of EuAuBi with FM configuration in P6$_{3}/mc$ space group. Red arrows indicate the direction of the magnetic moments. (c) The side view of the interlayer AFM EuAuBi. (d) Brillouin zone of bulk and the projected surface Brillouin zones of (100) planes.}
	\label{fig1}
\end{figure}

\section{METHODOLOGY }

Our first-principles calculations are carried out using density functional theory by the Vienna $ab$ $initio$ simulation package \cite{kresse1996efficiency,kresse1996efficient} with the projector augmented wave method \cite{blochl1994projector}. 
The energy cutoff for the plane-wave expansion is set as 400 eV.
The BZ integration is performed using a 9 × 9 × 5 $\Gamma$-centered $k$ mesh.
Perdew-Burke-Ernzerhof type of the exchange-correlation potential \cite{perdew1996generalized}, and Heyd-Scuseria-Ernzerhof (HSE) hybrid functional \cite{heyd2003hybrid} with mixing factor 0.25 are used to obtain the accurate electronic structures.
SOC is included in our calculations. Maximally localized Wannier functions (MLWFs) obtained from the WANNIER90 package \cite{mostofi2014updated}, is used to construct a tight-binding Hamiltonian. MLWFs has been symmetrized by the WannSymm package \cite{zhi2022wannsymm}. The topological properties including surface spectrum, Fermi arcs are calculated using the iterative Green’s function approach implemented in the WannierTools package \cite{wu2018wanniertools}. 

\begin{table*} [htbp]
	\centering
	\caption{Crystal structure, magnetism and calculated band characteristics of EuTX (T=Cu,Ag or Au and X= P, As, Sb, Bi). The band characteristics include direct band gap at the $\Gamma$ point $\delta_{\Gamma}$, energy difference of CBM between M and $\Gamma$ point $E_d$, and energy position (eV) of topological gap (TG) or Weyl point (WP). WS: Weyl semimetal; TSM: topological semimetal. }  
	\label{tab1} 
	\renewcommand\arraystretch{1.6}
	\begin{tabular}
		{cp{2.2cm}<{\centering}cp{2.2cm}<{\centering}cp{2.2cm}<{\centering}cp{2.2cm}<{\centering}cp{2.2cm}<{\centering}cp{2.2cm}<{\centering}cp{2.2cm}<{\centering}cp{2.2cm}<{\centering} } 
		\toprule  
		Compound  & \makecell{Space\\Group}& \makecell{Lattice\\ Constance ($\mathring{A}$)}& \makecell{Magnetic \\ Ordering} & \makecell{  $\delta_{\Gamma}$ (eV)}  & E$_d$ (eV)  & \makecell{Energy Position\\ (eV) of TG/WP} &\makecell{Topological\\ State} \\
		\midrule 	    
		EuCuP  &P6$_{3}$mmc& a=4.123, c=8.200 &FM    &   0.35 &   0.66  &                  &  Metal      \\
		EuAgP  &P6$_{3}$mmc& a=4.395, c=8.057 &FM    &   0.43 &   0.35  &                  &  Insulator  \\
		EuAuP  &P6$_{3}$mmc& a=4.313, c=8.258 &FM    &  -0.61 &  -0.02  & -0.016, 0.013    &  WS         \\
		EuCuAs &P6$_{3}$mmc& a=4.254, c=8.274 &A-AFM &   0.94 &   0.76  &                  &  Insulator  \\
		EuAgAs &P6$_{3}$mmc& a=4.516, c=8.107 &A-AFM &   0.33 &  -0.05  &                  &  Insulator  \\
		EuAuAs &P6$_{3}$mmc& a=4.445, c=8.285 &A-AFM &  -0.87 &  -0.78  & \makecell{-0.01 $\sim$ 0.02 \\ 0.002 $\sim$ 0.008 } &  TSM        \\
		EuCuSb &P6$_{3}$mmc& a=4.512, c=8.542 &A-AFM &   0.56 &   0.74  &                  &  Metal      \\
		EuAgSb &P6$_{3}$mmc& a=4.755, c=8.283 &A-AFM &   0.59 &   0.78  &                  &  Metal      \\
		EuAuSb &P6$_{3}$mmc& a=4.669, c=8.486 &A-AFM &  -0.45 &  -0.03  & \makecell{-0.033 $\sim$ -0.025 \\ -0.011 $\sim$ -0.008} &  TSM        \\
		EuCuBi &P6$_{3}$mmc& a=4.622, c=8.536 &A-AFM &  -0.35 &   0.34  & 0.467 $\sim $ 0.470 &  TSM        \\
		EuAgBi &P6$_{3}$mmc& a=4.877, c=8.139 &A-AFM &  -0.39 &   0.45  & 0.331 $\sim $ 0.336 &  TSM        \\ 
		EuAuBi &P6$_{3}$mc & a=4.799, c=8.295 &A-AFM &  -1.04 &  -0.43  &-0.07 $\sim $ -0.10 &  TSM        \\
		\bottomrule 
		\label{tab1}
	\end{tabular}
\end{table*}

\section{Crystal Structure and Symmetry}

All compounds in the EuTX family have structural properties that X and T form honeycomb lattice layers as shown in Fig. \ref{fig1}. X and T occupy the A and B sites in each honeycomb layer, respectively. These layers are stacked along the $z$-axis, with an alternation of the X and T atoms. The trigonal Eu layers are inserted between each X-T layer. 
In the most of compounds except EuAuBi, X and T atoms are in the same plane.
The crystal space group is hexagonal P6$_{3}/mmc$ (No.194), and the point group symmetry is $D_{6h}$.
Eu, X, and T sit at Wyckoff positions 2a (0,0,0), 2c (1/3,2/3,1/4), and 2d (1/3,2/3,3/4) as shown in Fig. \ref{fig1}(a), respectively. 
The P6$_{3}/mmc$ group contains the following generators: threefold rotation $C_{3z}$, inversion $\mathcal{P}$, twofold screw rotation $S_{2z}=\{C_{2z}|00\frac{1}{2}\}$ and twofold rotation $C_{2y}$. 
Unlike other compounds, Au and Bi atoms in EuAuBi slightly deviate from the original planar honeycomb layers in Fig. \ref{fig1}(b). Au atoms move upward and Bi atoms move downward, which leads to the absence of inversion center.
The resulted polar hexagonal structure is in the space group P6$_{3}/mc$ (No.186), in which Eu is at the Wyckoff position of $2a$ (0,0,0), Au and Bi are at the position of 2b (1/3,2/3,z$_{t}$) and (1/3,2/3,z$_{x}$).
Due to the broken inversion symmetry, the point group symmetry is reduced from  $D_{6h}$ to $C_{6v}$.
The experimental lattice constants of EuTX family \cite{tomuschat1981abx,merlo1990rmx} are used in our calculations, and listed in Table \ref{tab1}. 
The ionic relaxation is performed for EuAuBi, and the optimized coordinates z$_{t}$=0.311, z$_{x}$=0.720, which agree well with experimental results z$_{t}$=0.294, z$_{x}$=0.730, are used in the following calculations.

Experimentally, EuTP shows a low temperature ferromagnetic behavior, while EuTAs/Sb/Bi show antiferromagnetic behavior \cite{tomuschat1984magnetic}.
The direction of magnetic moments in FM EuCuP prefers to $z$-axis \cite{iha2019anomalous}.
The antiferromagnetic structure of EuTX family is always A-type AFM order, in which intralayer Eu$^{2+}$ is FM order and interlayer Eu$^{2+}$ is AFM order.
However, the direction of the magnetic moments still has some ambiguity. 
By magnetic field dependence of isothermal magnetization, the sharper magnetization curve for $H//ab$ than $H//c$ indicates the magnetic moments lie in the plane in EuCuAs, EuCuSb, EuAgAs, EuCuBi, and EuAuBi \cite{tong2014magnetic,takahashi2020competing,laha2021topological,wang2023structure,takahashi2023superconductivity,may2023coupling}.
Here, we assume that all the magnetic moments of FM states are along the $z$-axis and all the magnetic moments of AFM states are along the $x$-axis unless specifically indicated in our calculation.

\section{RESULT}

\subsection{A. Electronic structures in EuTX family}
\begin{figure*}[htbp]
	\centering
	\includegraphics[width=1\textwidth]{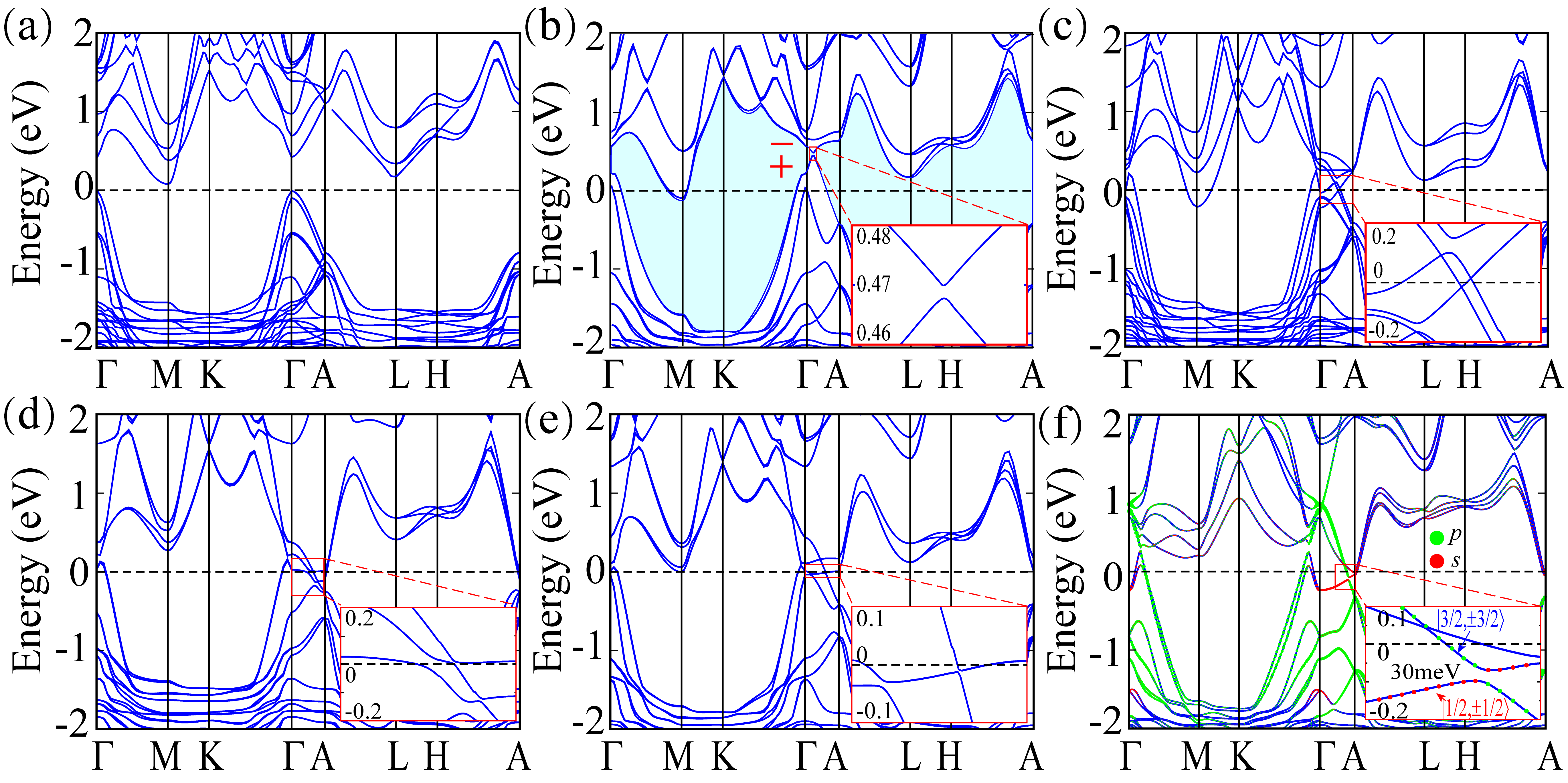}
	\caption{The HSE calculated band structures of (a) EuAgP, (b) EuCuBi, (c)EuAuP, (d)EuAuAs, (e)EuAuSb, and (f)EuAuBi. A continuous band gap around the Fermi level in (b) is marked by the blue shade. The insets in (b-f) are the zoom-in of the band structures. The red and green dots in (f) represent the projections of the Au $s$ and Bi $p$ orbitals, respectively.}
	\label{fig2}
\end{figure*}

We calculate the electronic structures of 12 compounds in the EuTX family. In order to find new topological materials, we count several band characteristics that can determine band structures and topological features, and list them in Table \ref{tab1}. 
The band structures of EuTX family have features that the valance band maximum (VBM) always occurs at the $\Gamma$ point, while the conduction band minimum (CBM) occurs at the $\Gamma$ or M point. 
By comparing $E_d=E_{CBM,\Gamma}-E_{CBM,M}$ in Table \ref{tab1}, it shows that CBM of EuAuX (X=P, As, Sb, and Bi) and EuAgAs occur at the $\Gamma$ point and CBM of other compounds occur at M point.  
To realize topological states, it needs to happen band inversion at $\Gamma$ point, which means the value of $\delta_{\Gamma}=E_{CBM,\Gamma}-E_{VBM,\Gamma}<0$ in Table \ref{tab1}.
In our calculations, EuAgP, EuCuAs, and EuAgAs are trivial insulators with the band gap of 0.08 eV, 0.18 eV, and 0.33 eV, respectively. 
EuCuP, EuCuSb, and EuAgSb are topological trivial metals and other materials are topological nontrivial.
Particularly, the band structures of EuAgP plotted in Fig. \ref{fig2}(a) is inconsistent with the previous reports that FM EuAgP is a Weyl semimetal \cite{ge2022ferromagnetic}.
This is because HSE exchange-correlation functional is used in our calculation, which is believed more accurate for the band gap calculation than GGA functional \cite{barman2020symmetry}.

In our calculations, we find $\delta_{\Gamma}<0$ and $E_{d}>0$ in EuCuBi and EuAgBi, which means it happens band inversion at $\Gamma$ point and exists electron pocket around the $M$ point. 
Since EuCuBi and EuAgBi have similar band structures, we only plot the band structures of EuCuBi in Fig. \ref{fig2}(b) as representation, which manifests that the band crossing along $\Gamma$-A is above the Fermi level about 0.46 eV in EuCuBi (0.33 eV above Fermi level in EuAgBi).
Considering the AFM configuration with in-plane magnetic moments as shown in Fig. \ref{fig1}(c), SOC could open a topological nontrivial gap at the band crossing as shown by the inset of Fig. \ref{fig2}(b).
Given that EuCuBi possesses inversion symmetry and a continuous band gap (blue region in Fig. \ref{fig2}(b)), its nontrivial topology can be characterized by the parity-based invariant $\mathbb{Z}_{4}$, which is defined as  \cite{turner2012quantized,watanabe2018structure,ono2018unified},
\begin{equation}
\mathbb{Z}_{4}=\sum_{\alpha=1}^{8}\sum_{n=1}^{n_{occ}}\frac{1+\xi_{n}(\Lambda_{\alpha})}{2} \quad mod \quad 4,
\end{equation}
where $\xi_{n}(\Lambda_{\alpha})$ is the parity eigenvalue of the $n$th band at eight time-invariant momenta $\Lambda_{\alpha}$, and $n_{occ}$ is the total number of valence electrons.
As shown in Fig. \ref{fig2}(b), the two inverted bands at $\Gamma$ point have opposite parity. Our calculated results give $\mathbb{Z}_{4}=2$, confirming its nontrivial topological property.

For ideal topological semimetals in the EuTX family, it needs $\delta_{\Gamma}<0$ and $E_d<0$, which means the CBM at M point is away from the Fermi level.
Meanwhile, it leads the topological gap or nodes just located the Fermi level to dominate the low-energy physics.
In our calculations, we find that EuAuX (X = P, As, Sb, and Bi) are ideal topological semimetals.
As shown in Fig. \ref{fig2}(c), EuAuP is Weyl semimetal due to FM configuration, which hosts multiple Weyl points near the Fermi level along $\Gamma$-A.
Especially, the energy of two Weyl points is located at $k_z=0.275$ (2$\pi$/c) and $k_z=0.300$ (2$\pi$/c) is at -0.016 eV and 0.013 eV, respectively (inset of Fig. \ref{fig2}(c)).
On the other hand, EuAuAs, EuAuSb, and EuAuBi are AFM semimetals with topological nontrivial gap near the Fermi level as shown in Fig. \ref{fig2}(d-f).
For EuAuAs and EuAuSb, there are two topological gaps along $\Gamma$-A, which are located at -0.01 $\sim$ 0.02 eV, 0.002 $\sim$ 0.008 eV in EuAuAs (inset of Fig. \ref{fig2}(d)) and -0.033 $\sim$ -0.025 eV, -0.011 $\sim$ -0.008 eV in EuAuSb (inset of Fig. \ref{fig2}(e)).
Particularly for EuAuBi, there is only one topological gap along $\Gamma$-A at -0.10 $\sim$ -0.07 eV, which has the largest topological band gap among the EuTX family.
Therefore, We will discuss electronic structures and topological properties of EuAuBi further in section B.

\subsection{B. Topological feature in EuAuBi}

We calculate and list the magnetic space group, energy and magnetic moments on each Eu atom of different magnetic configurations of EuAuBi in Table \ref{tab2}. 
The results demonstrate that all magnetic states are lower than the nonmagnetic states about 7.3 eV/f.u. and the energy of A-type AFM configuration is lower than ferromagnetic configuration about 2.0 meV/f.u., which indicates interlayer coupling prefers AFM configuration.
The magnetic ground state of EuAuBi is A-type AFM configuration with 6.86 $\mu_B$ magnetic moments along $x$-axis (AFM-$x$), which agrees with the experiment well \cite{takahashi2023superconductivity}. 
Due to the energy of A-type AFM states with $z$ moments (AFM-$z$) is just 0.2 meV higher than the AFM-$x$ ground state for each formula, the magnetic moments can be switched to $z$-axis by magnetic field.
Therefore, we will discuss the electronic structures of both the AFM-$x$ and AFM-$z$ states.

\begin{table*}[htb]
	\centering 
	\caption{Comparison of different magnetic configuration for EuAuBi. Here, we list MSG, total energies calculated by GGA+SOC+U with the value of U = 7eV, and the converged magnetic moments of each Eu atom.}  
	\label{tab2} 
	\renewcommand\arraystretch{1.5}
	\begin{tabular}{cp{3cm}<{\centering}cp{3cm}<{\centering}cp{3cm}<{\centering}cp{3cm}<{\centering}cp{3cm}<{\centering}} 
		\toprule  
		Config.      & MSG   & Eu$_{1} (\mu_B)$   & Eu$_{2} (\mu_B)$    & Energy (eV/f.u.)   \\
		\midrule 
		Nonmagnetic & $P6_3mc$   & (0,0,0)     &(0,0,0)        & -13.1478   \\
		FM-$z$         & $P6_3m'c'$          & (0,0,6.86) & (0,0,6.86)  & -20.4519              \\
		FM-$x$         & $Pmc'2'$          & (6.86,0,0) & (6.86,0,0)  & -20.4520              \\
		AFM-$z$  & $P6_{3}'m'c$ & (0,0,6.86) & (0,0,-6.86)    & -20.4540        \\
		AFM-$x$   & $Pmc2$ & (6.86,0,0)       & (-6.86,0,0)           & -20.4542       \\

		\bottomrule 
	\end{tabular}
\end{table*}

\begin{figure*}[htb]
	\centering
	\includegraphics[width=1\textwidth]{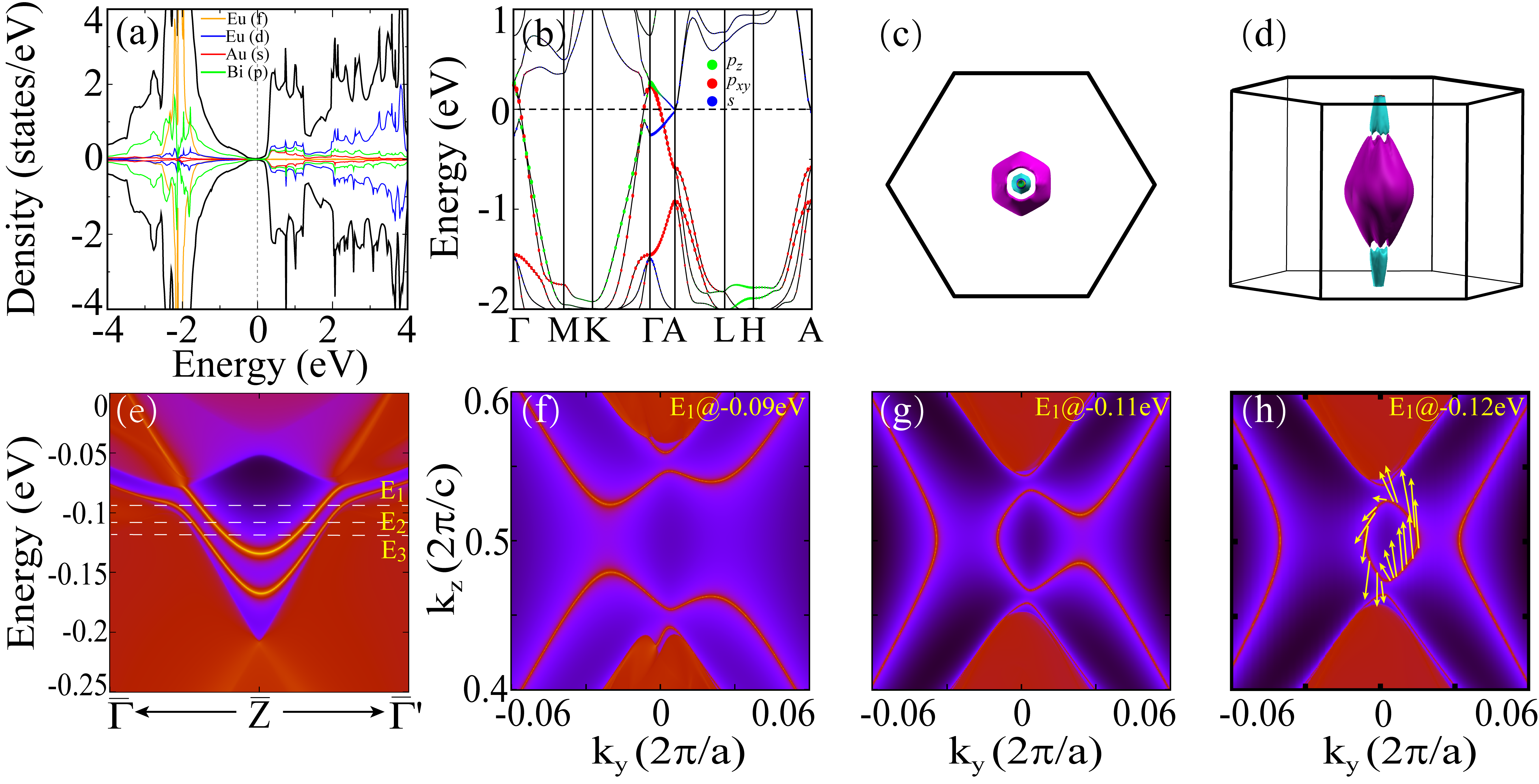}
	\caption{ The electronic structures of AFM-$x$ EuAuBi. (a) The Total and Projected density of states without SOC. (b) The band structures with spectral weight of Bi-$p_{z}$ (green), Bi-$p_{xy}$ (red), and Au-$s$ (blue) orbitals without SOC. (c-d) Top and side view of the bulk Fermi surface in the Brillouin zone. (e) Energy and momentum dependence of the LDOS on (100) surface. (f-h) The Fermi surface on (100) the surface with chemical potential at -0.09 eV, -0.11 eV, and -0.12 eV, respectively. The yellow arrows in (h) show a right-handed spin helical texture. }
	\label{fig3}
\end{figure*}

The projected density of states of AFM EuAuBi without SOC are shown in Fig. \ref{fig3}(a). 
The analysis of orbital character shows that the half-filled $4f$ orbitals of Eu$^{2+}$ ions are pushed down to 1.0 - 2.5 eV below the Fermi level, while the unoccupied $4f$ orbitals are pushed above the Fermi level more than 4 eV.
The states near the Fermi level are mainly contributed from the $s$ orbitals of Au atoms, and $p$ orbitals of Bi atoms hybridized with considerable $d$ orbitals of Eu atoms. 
As shown in Fig. \ref{fig3}(b), the $p$ orbitals invert with the $s$ orbital at $\Gamma$ point, and band crossing between $p_{xy}$ and $s$ orbital exists obviously along $\Gamma$-A when SOC is excluded.
By including the SOC effect, the new eigenstates can be written as $|J,J_{z}\rangle$, where $J$ means total angular momentum and $J_{z}$ means $z$ projection of total angular momentum.
$s$ orbitals form two states $|\frac{1}{2},\pm\frac{1}{2}\rangle_{s}$, and $p$ orbitals form six states $|\frac{3}{2},\pm\frac{3}{2}\rangle_{p}$, $|\frac{3}{2},\pm\frac{1}{2}\rangle_{p}$, $|\frac{1}{2},\pm\frac{1}{2}\rangle_{p}$ .
Due to the SOC effect, $|\frac{3}{2},\pm\frac{3}{2}\rangle_{p}$ states are pushed up, while the energy of $|\frac{3}{2},\pm\frac{1}{2}\rangle_{p}$ and $|\frac{1}{2},\pm\frac{1}{2}\rangle_{p}$ states are lower.
Finally, the low-energy physics are dominated by $|\frac{3}{2},\pm\frac{3}{2}\rangle_{p}$ and $|\frac{1}{2},\pm\frac{1}{2}\rangle_{s}$ states as manifested by inset of Fig. \ref{fig2}(f).  
As a result, two $p$ orbitals go upward to cross the Fermi level and two $s$ orbitals go downward, which forms two thin torus-shaped hole pockets (purple pockets) around $\Gamma$ points and two electric pockets (blue pockets) around A points as shown in Fig. \ref{fig3}(c-d).

When the magnetic moments are along $x$-axis, the system is in type-I MSG $Pmc2$.
The system breaks threefold rotation symmetry $C_{3z}$ and only preserves twofold screw rotation $S_{2z}=\{C_{2z}|00\frac{1}{2}\}$ and two vertical mirror symmetry  $M_{x}$ and $\tilde{M}_{y}=\{M_{y}|00\frac{1}{2}\}$. 
The corresponding little group of $k$ points on $\Gamma$-A is $C_{2v}$ containing $C_{2z}$ and $M_{x}$.
In the $|J_{z}=\pm 3/2\rangle $ (or $|J_{z}=\pm 1/2\rangle$) basis, we can write $C_{2z}=i\sigma_{z}$ , $M_{x}=i\sigma_{x}$, the non-commutation relation $[C_{2z},M_{x}] \neq 0$ enforces two-dimensional irreducible representations, which leads doubly degenerate band along $\Gamma$-A.
Meanwhile, $C_{2z}$ cannot prohibit the hybridization between the different $|J_{z}|$ eigenstates, because  the representation of $|J_{z}=\pm 3/2\rangle $ and $|J_{z}=\pm 1/2\rangle$ is same. 
Thus it opens a 30 meV topological gap between $|J_{z}=\pm 3/2\rangle $ and $|J_{z}=\pm 1/2\rangle$ as shown in the inset of Fig. \ref{fig2}(f).
Because a continuous band gap is maintained in the whole BZ, the occupied states of EuAuBi are equal to  topological insulator like we mentioned in AFM topological material EuCuBi . 
The topological surface states are expected to exist between the topological gap.

\begin{figure*}[htb]
	\centering
	\includegraphics[width=1\textwidth]{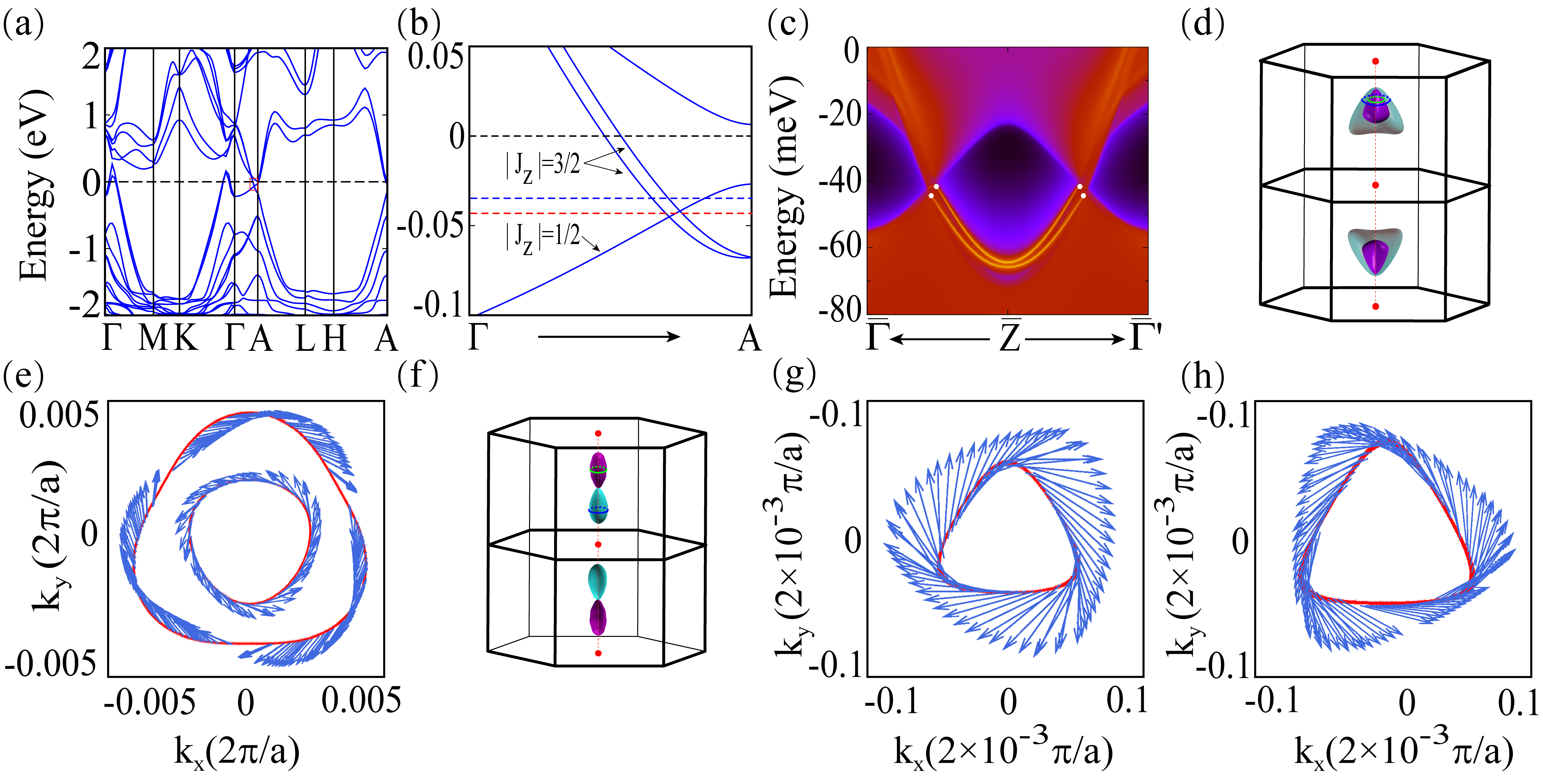}
	\caption{ The electronic structures of AFM-$z$ EuAuBi. (a) Band structures of AFM-$z$ EuAuBi with SOC. (b) Zoom-in of the band structures near A point to clearly show the TDNPs. (c) The LDOS on (100) surface of AFM-$z$ EuAuBi. (d)  The calculated Fermi surfaces near the TDNPs with chemical potential at -30 meV, where the Fermi surfaces are magnified 50 times to visibly exhibit them and their tangency. (e) The horizontal spin texture of Fermi surface in (d) at $k_{z}=0.48$. (f) The 100 times magnified Fermi surfaces near the TDNPs when chemical potential sits between the TDNPs. (g-h) The horizontal spin texture of hole pocket and electron pocket in (f), respectively.
	}
	\label{fig4}
\end{figure*}

We calculate surface states based on Green’s functions of the semi-infinite system, which are constructed by the MLWFs. 
Due to the breaking of time reversal symmetry, the surface local density of states (LDOS) on (100) surface do not degenerate at $\bar{Z}$ point, which gives rise to a gapped topological state as shown in  Fig. \ref{fig3}(e) . 
For the convenience of experimental observation like angle-resolved photoemission spectroscopy \cite{lv2019angle}, the evolution of Fermi surface with the different chemical potentials (dashed line in Fig. \ref{fig3}(e)) are shown in Fig. \ref{fig3}(f-h).

Next, we consider EuAuBi in the AFM-$z$ states, whose band structures are plotted in Fig. \ref{fig4}(a). 
The system changes to type-\uppercase\expandafter{\romannumeral3} MSG $P6'_{3}m'c$ and little group of $k$ points on $\Gamma$-A is $C_{3v}$ group, which consist of $C_{3z}$ and $M_{x}$. 
For $|J_{z}=\pm1/2\rangle$ subspace, $C_{3z}=e^{-i\frac{\pi}{3}\sigma_{z}}$ and $M_x=-i\sigma_{x}$, the non-commutation relation of $C_{3z}$ and $M_x$ still enforces the doubly degenerate bands.
However for $|J_{z}=\pm3/2\rangle$ subspace,  $C_{3z}=-\sigma_{0}$ and $M_{x}=i\sigma_{x}$, the commutator of $[C_{3z},M_{x}]=0$ leads two one-dimensional irreducible representations.
Furthermore, $C_{3z}$ symmetry prohibits different $|J_{z}|$ eigenstates hybridized with each other. 
Therefore, when there is a band inversion at $\Gamma$ point, it leads to two protected TDNPs along $\Gamma$-A as shown in Fig. \ref{fig4}(b). 

\begin{figure}[htb]
	\centering
	\includegraphics[width=0.5\textwidth]{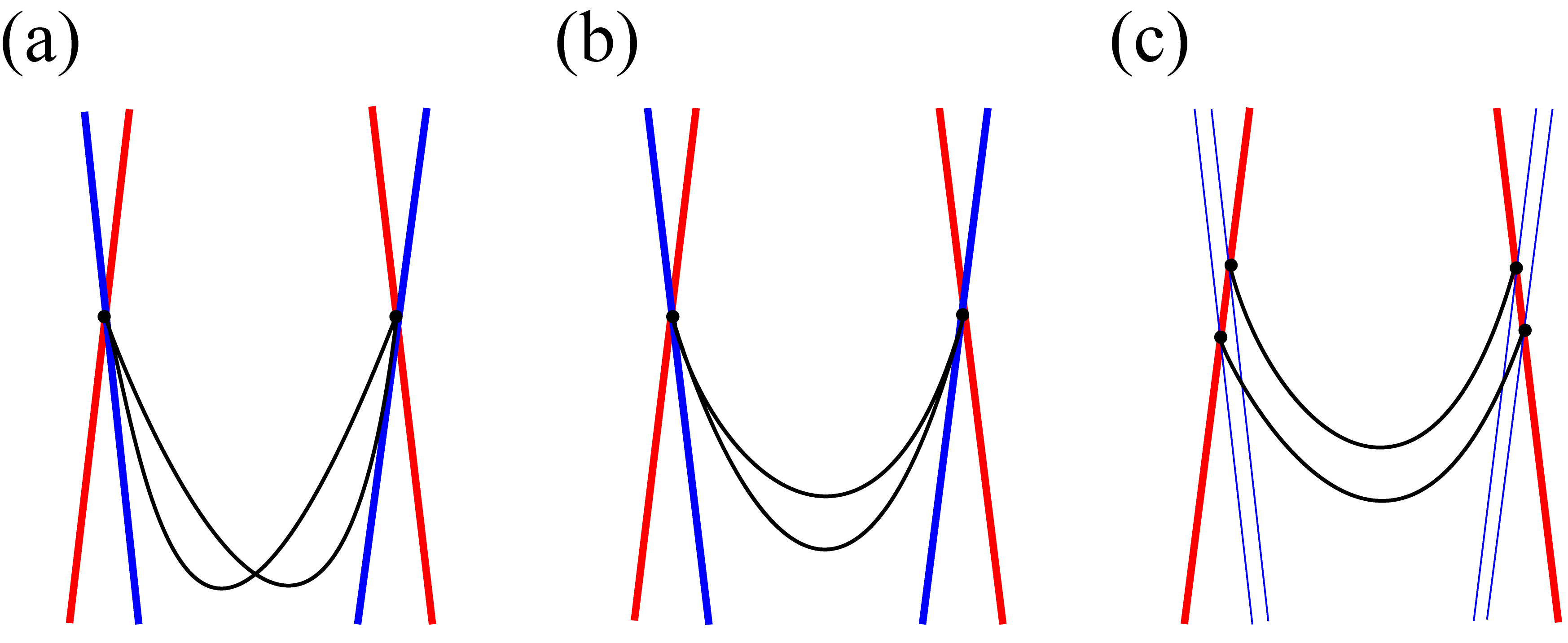}
	\caption{ Schematic of the topological surface states evolution from nonmagnetic topological Dirac semimetal (a), to magnetic topological Dirac semimetal (b), and to magnetic topological TDNP semimetal (c).
	}
	\label{fig5}
\end{figure}

The magnetic topological TDNP semimetal has following three characteristics.
The most remarkable characteristic is a pair of separated surface Fermi arcs connecting the TDNPs from the bulk states as shown in Fig. \ref{fig4}(c).
We can illustrate the origin of the surface states as shown in Fig. \ref{fig5}. 
Starting from nonmagnetic topological Dirac semimetal in Fig. \ref{fig5}(a), there is a pair of Dirac cone like surface states on each surface. 
If the bulk Dirac points can still be protected when introducing the magnetic ordering, it is called magnetic Dirac semimetal.
For magnetic topological Dirac semimetal, the surface Dirac cone will open a gap like Fig. \ref{fig5}(b) \cite{ma2020emergence}. 
When one of the doubly degenerate band splits like $|J_{z}=\pm3/2\rangle$ states in AFM-$z$ EuAuBi, the bulk Dirac point splits into a pair of TDNPs as shown in Fig. \ref{fig5}(c).
Due to the splitting of the bulk Dirac point, two branches of the Fermi arcs are fully separated like Fig. \ref{fig4}(c).

Secondly, for any chemical potential near the TDNPs, there are always two carrier pockets that are tangent to each other at some point along $\Gamma$-A, and each pocket contains one TNDP.
For instance, when we set chemical potential at -0.03 eV sitting above the TDNPs (blue dashed line in Fig.  \ref{fig4}(b)),  there are two electron pockets that are internally tangent to each other (Fig. \ref{fig4}(d)). 
When chemical potential sits between the TDNPs (red dashed line in Fig. \ref{fig4}(b)), there is one electron and one hole pocket that is externally tangent (Fig. \ref{fig4}(f)).

Finally, along any horizontal loop on the Fermi surface near the TDNPs, the spin winds exactly one round around the $z$-axis, and the two tangent Fermi surface have opposite winding numbers.  
This is a topological robust feature of the Fermi surface near the TDNPs.
For example, we calculate the horizontal spin texture on Fermi surface in Fig. \ref{fig4}(d) at k$_z$=0.48 $(2\pi/c)$, and show the two tangent Fermi surface hosting opposite winding numbers in Fig. \ref{fig4}(e).
We also calculate the spin texture on each Fermi pocket in Fig. \ref{fig4}(f), and see two tangent electron pocket and hole pocket having opposite winding numbers as shown in Fig. \ref{fig4}(g-h). 

\section{CONCLUSION}

In conclusion, based on first-principles calculations, we find multiple topological materials in EuTX (T=Cu, Ag, Au; X=P, As, Sb, Bi) family. 
Particularly, EuAuP is FM Weyl semimetal with two Weyl points along $\Gamma$-A within 20 meV of the Fermi level, and EuAuX (X= As, Sb, and Bi) in their ground state with in-plane magnetic moments are AFM semimetals with topological nontrivial gap near the Fermi level. 
Moreover, many different types of topological states can be realized in EuAuX by tuning the magnetic moments.
For example, when changing the magnetic moments from in-plane to $z$-axis , EuAuX (X= As, Sb, and Bi) evolve into topological TDNP semimetals.
We also discuss the main characteristics of TDNPs in AFM-$z$ EuAuBi, including the surface Fermi arcs, and tangent Fermi surfaces with opposite spin winding numbers.
Our results provide a platform for studying the magnetic topological insulating states and other exotic magnetic topological states.

\section{ACKNOWLEDGMENTS}

This work is supported by the National Natural Science Foundation of China (12274154).

\bibliography{ref}

\begin{thebibliography}{70}%
\makeatletter
\providecommand \@ifxundefined [1]{%
 \@ifx{#1\undefined}
}%
\providecommand \@ifnum [1]{%
 \ifnum #1\expandafter \@firstoftwo
 \else \expandafter \@secondoftwo
 \fi
}%
\providecommand \@ifx [1]{%
 \ifx #1\expandafter \@firstoftwo
 \else \expandafter \@secondoftwo
 \fi
}%
\providecommand \natexlab [1]{#1}%
\providecommand \enquote  [1]{``#1''}%
\providecommand \bibnamefont  [1]{#1}%
\providecommand \bibfnamefont [1]{#1}%
\providecommand \citenamefont [1]{#1}%
\providecommand \href@noop [0]{\@secondoftwo}%
\providecommand \href [0]{\begingroup \@sanitize@url \@href}%
\providecommand \@href[1]{\@@startlink{#1}\@@href}%
\providecommand \@@href[1]{\endgroup#1\@@endlink}%
\providecommand \@sanitize@url [0]{\catcode `\\12\catcode `\$12\catcode
  `\&12\catcode `\#12\catcode `\^12\catcode `\_12\catcode `\%12\relax}%
\providecommand \@@startlink[1]{}%
\providecommand \@@endlink[0]{}%
\providecommand \url  [0]{\begingroup\@sanitize@url \@url }%
\providecommand \@url [1]{\endgroup\@href {#1}{\urlprefix }}%
\providecommand \urlprefix  [0]{URL }%
\providecommand \Eprint [0]{\href }%
\providecommand \doibase [0]{https://doi.org/}%
\providecommand \selectlanguage [0]{\@gobble}%
\providecommand \bibinfo  [0]{\@secondoftwo}%
\providecommand \bibfield  [0]{\@secondoftwo}%
\providecommand \translation [1]{[#1]}%
\providecommand \BibitemOpen [0]{}%
\providecommand \bibitemStop [0]{}%
\providecommand \bibitemNoStop [0]{.\EOS\space}%
\providecommand \EOS [0]{\spacefactor3000\relax}%
\providecommand \BibitemShut  [1]{\csname bibitem#1\endcsname}%
\let\auto@bib@innerbib\@empty
\bibitem [{\citenamefont {Hasan}\ and\ \citenamefont
  {Kane}(2010)}]{hasan2010colloquium}%
  \BibitemOpen
  \bibfield  {author} {\bibinfo {author} {\bibfnamefont {M.~Z.}\ \bibnamefont
  {Hasan}}\ and\ \bibinfo {author} {\bibfnamefont {C.~L.}\ \bibnamefont
  {Kane}},\ }\href@noop {} {\bibfield  {journal} {\bibinfo  {journal} {Reviews
  of modern physics}\ }\textbf {\bibinfo {volume} {82}},\ \bibinfo {pages}
  {3045} (\bibinfo {year} {2010})}\BibitemShut {NoStop}%
\bibitem [{\citenamefont {Qi}\ and\ \citenamefont
  {Zhang}(2011)}]{qi2011topological}%
  \BibitemOpen
  \bibfield  {author} {\bibinfo {author} {\bibfnamefont {X.-L.}\ \bibnamefont
  {Qi}}\ and\ \bibinfo {author} {\bibfnamefont {S.-C.}\ \bibnamefont {Zhang}},\
  }\href@noop {} {\bibfield  {journal} {\bibinfo  {journal} {Reviews of Modern
  Physics}\ }\textbf {\bibinfo {volume} {83}},\ \bibinfo {pages} {1057}
  (\bibinfo {year} {2011})}\BibitemShut {NoStop}%
\bibitem [{\citenamefont {Chiu}\ \emph {et~al.}(2016)\citenamefont {Chiu},
  \citenamefont {Teo}, \citenamefont {Schnyder},\ and\ \citenamefont
  {Ryu}}]{chiu2016classification}%
  \BibitemOpen
  \bibfield  {author} {\bibinfo {author} {\bibfnamefont {C.-K.}\ \bibnamefont
  {Chiu}}, \bibinfo {author} {\bibfnamefont {J.~C.}\ \bibnamefont {Teo}},
  \bibinfo {author} {\bibfnamefont {A.~P.}\ \bibnamefont {Schnyder}},\ and\
  \bibinfo {author} {\bibfnamefont {S.}~\bibnamefont {Ryu}},\ }\href@noop {}
  {\bibfield  {journal} {\bibinfo  {journal} {Reviews of Modern Physics}\
  }\textbf {\bibinfo {volume} {88}},\ \bibinfo {pages} {035005} (\bibinfo
  {year} {2016})}\BibitemShut {NoStop}%
\bibitem [{\citenamefont {Armitage}\ \emph {et~al.}(2018)\citenamefont
  {Armitage}, \citenamefont {Mele},\ and\ \citenamefont
  {Vishwanath}}]{armitage2018weyl}%
  \BibitemOpen
  \bibfield  {author} {\bibinfo {author} {\bibfnamefont {N.}~\bibnamefont
  {Armitage}}, \bibinfo {author} {\bibfnamefont {E.}~\bibnamefont {Mele}},\
  and\ \bibinfo {author} {\bibfnamefont {A.}~\bibnamefont {Vishwanath}},\
  }\href@noop {} {\bibfield  {journal} {\bibinfo  {journal} {Reviews of Modern
  Physics}\ }\textbf {\bibinfo {volume} {90}},\ \bibinfo {pages} {015001}
  (\bibinfo {year} {2018})}\BibitemShut {NoStop}%
\bibitem [{\citenamefont {Kane}\ and\ \citenamefont
  {Mele}(2005)}]{kane2005quantum}%
  \BibitemOpen
  \bibfield  {author} {\bibinfo {author} {\bibfnamefont {C.~L.}\ \bibnamefont
  {Kane}}\ and\ \bibinfo {author} {\bibfnamefont {E.~J.}\ \bibnamefont
  {Mele}},\ }\href@noop {} {\bibfield  {journal} {\bibinfo  {journal} {Physical
  review letters}\ }\textbf {\bibinfo {volume} {95}},\ \bibinfo {pages}
  {226801} (\bibinfo {year} {2005})}\BibitemShut {NoStop}%
\bibitem [{\citenamefont {Bernevig}\ \emph {et~al.}(2006)\citenamefont
  {Bernevig}, \citenamefont {Hughes},\ and\ \citenamefont
  {Zhang}}]{bernevig2006quantum}%
  \BibitemOpen
  \bibfield  {author} {\bibinfo {author} {\bibfnamefont {B.~A.}\ \bibnamefont
  {Bernevig}}, \bibinfo {author} {\bibfnamefont {T.~L.}\ \bibnamefont
  {Hughes}},\ and\ \bibinfo {author} {\bibfnamefont {S.-C.}\ \bibnamefont
  {Zhang}},\ }\href@noop {} {\bibfield  {journal} {\bibinfo  {journal}
  {science}\ }\textbf {\bibinfo {volume} {314}},\ \bibinfo {pages} {1757}
  (\bibinfo {year} {2006})}\BibitemShut {NoStop}%
\bibitem [{\citenamefont {Teo}\ \emph {et~al.}(2008)\citenamefont {Teo},
  \citenamefont {Fu},\ and\ \citenamefont {Kane}}]{teo2008surface}%
  \BibitemOpen
  \bibfield  {author} {\bibinfo {author} {\bibfnamefont {J.~C.}\ \bibnamefont
  {Teo}}, \bibinfo {author} {\bibfnamefont {L.}~\bibnamefont {Fu}},\ and\
  \bibinfo {author} {\bibfnamefont {C.}~\bibnamefont {Kane}},\ }\href@noop {}
  {\bibfield  {journal} {\bibinfo  {journal} {Physical Review B}\ }\textbf
  {\bibinfo {volume} {78}},\ \bibinfo {pages} {045426} (\bibinfo {year}
  {2008})}\BibitemShut {NoStop}%
\bibitem [{\citenamefont {Zhang}\ \emph {et~al.}(2009)\citenamefont {Zhang},
  \citenamefont {Liu}, \citenamefont {Qi}, \citenamefont {Dai}, \citenamefont
  {Fang},\ and\ \citenamefont {Zhang}}]{zhang2009topological}%
  \BibitemOpen
  \bibfield  {author} {\bibinfo {author} {\bibfnamefont {H.}~\bibnamefont
  {Zhang}}, \bibinfo {author} {\bibfnamefont {C.-X.}\ \bibnamefont {Liu}},
  \bibinfo {author} {\bibfnamefont {X.-L.}\ \bibnamefont {Qi}}, \bibinfo
  {author} {\bibfnamefont {X.}~\bibnamefont {Dai}}, \bibinfo {author}
  {\bibfnamefont {Z.}~\bibnamefont {Fang}},\ and\ \bibinfo {author}
  {\bibfnamefont {S.-C.}\ \bibnamefont {Zhang}},\ }\href@noop {} {\bibfield
  {journal} {\bibinfo  {journal} {Nature physics}\ }\textbf {\bibinfo {volume}
  {5}},\ \bibinfo {pages} {438} (\bibinfo {year} {2009})}\BibitemShut {NoStop}%
\bibitem [{\citenamefont {Yan}\ \emph {et~al.}(2010)\citenamefont {Yan},
  \citenamefont {Liu}, \citenamefont {Zhang}, \citenamefont {Yam},
  \citenamefont {Qi}, \citenamefont {Frauenheim},\ and\ \citenamefont
  {Zhang}}]{yan2010theoretical}%
  \BibitemOpen
  \bibfield  {author} {\bibinfo {author} {\bibfnamefont {B.}~\bibnamefont
  {Yan}}, \bibinfo {author} {\bibfnamefont {C.-X.}\ \bibnamefont {Liu}},
  \bibinfo {author} {\bibfnamefont {H.-J.}\ \bibnamefont {Zhang}}, \bibinfo
  {author} {\bibfnamefont {C.-Y.}\ \bibnamefont {Yam}}, \bibinfo {author}
  {\bibfnamefont {X.-L.}\ \bibnamefont {Qi}}, \bibinfo {author} {\bibfnamefont
  {T.}~\bibnamefont {Frauenheim}},\ and\ \bibinfo {author} {\bibfnamefont
  {S.-C.}\ \bibnamefont {Zhang}},\ }\href@noop {} {\bibfield  {journal}
  {\bibinfo  {journal} {Europhysics Letters}\ }\textbf {\bibinfo {volume}
  {90}},\ \bibinfo {pages} {37002} (\bibinfo {year} {2010})}\BibitemShut
  {NoStop}%
\bibitem [{\citenamefont {Wan}\ \emph {et~al.}(2011)\citenamefont {Wan},
  \citenamefont {Turner}, \citenamefont {Vishwanath},\ and\ \citenamefont
  {Savrasov}}]{wan2011topological}%
  \BibitemOpen
  \bibfield  {author} {\bibinfo {author} {\bibfnamefont {X.}~\bibnamefont
  {Wan}}, \bibinfo {author} {\bibfnamefont {A.~M.}\ \bibnamefont {Turner}},
  \bibinfo {author} {\bibfnamefont {A.}~\bibnamefont {Vishwanath}},\ and\
  \bibinfo {author} {\bibfnamefont {S.~Y.}\ \bibnamefont {Savrasov}},\
  }\href@noop {} {\bibfield  {journal} {\bibinfo  {journal} {Physical Review
  B}\ }\textbf {\bibinfo {volume} {83}},\ \bibinfo {pages} {205101} (\bibinfo
  {year} {2011})}\BibitemShut {NoStop}%
\bibitem [{\citenamefont {Xu}\ \emph {et~al.}(2011)\citenamefont {Xu},
  \citenamefont {Weng}, \citenamefont {Wang}, \citenamefont {Dai},\ and\
  \citenamefont {Fang}}]{xu2011chern}%
  \BibitemOpen
  \bibfield  {author} {\bibinfo {author} {\bibfnamefont {G.}~\bibnamefont
  {Xu}}, \bibinfo {author} {\bibfnamefont {H.}~\bibnamefont {Weng}}, \bibinfo
  {author} {\bibfnamefont {Z.}~\bibnamefont {Wang}}, \bibinfo {author}
  {\bibfnamefont {X.}~\bibnamefont {Dai}},\ and\ \bibinfo {author}
  {\bibfnamefont {Z.}~\bibnamefont {Fang}},\ }\href@noop {} {\bibfield
  {journal} {\bibinfo  {journal} {Physical review letters}\ }\textbf {\bibinfo
  {volume} {107}},\ \bibinfo {pages} {186806} (\bibinfo {year}
  {2011})}\BibitemShut {NoStop}%
\bibitem [{\citenamefont {Burkov}\ and\ \citenamefont
  {Balents}(2011)}]{burkov2011weyl}%
  \BibitemOpen
  \bibfield  {author} {\bibinfo {author} {\bibfnamefont {A.}~\bibnamefont
  {Burkov}}\ and\ \bibinfo {author} {\bibfnamefont {L.}~\bibnamefont
  {Balents}},\ }\href@noop {} {\bibfield  {journal} {\bibinfo  {journal}
  {Physical review letters}\ }\textbf {\bibinfo {volume} {107}},\ \bibinfo
  {pages} {127205} (\bibinfo {year} {2011})}\BibitemShut {NoStop}%
\bibitem [{\citenamefont {Zyuzin}\ \emph {et~al.}(2012)\citenamefont {Zyuzin},
  \citenamefont {Wu},\ and\ \citenamefont {Burkov}}]{zyuzin2012weyl}%
  \BibitemOpen
  \bibfield  {author} {\bibinfo {author} {\bibfnamefont {A.}~\bibnamefont
  {Zyuzin}}, \bibinfo {author} {\bibfnamefont {S.}~\bibnamefont {Wu}},\ and\
  \bibinfo {author} {\bibfnamefont {A.}~\bibnamefont {Burkov}},\ }\href@noop {}
  {\bibfield  {journal} {\bibinfo  {journal} {Physical Review B}\ }\textbf
  {\bibinfo {volume} {85}},\ \bibinfo {pages} {165110} (\bibinfo {year}
  {2012})}\BibitemShut {NoStop}%
\bibitem [{\citenamefont {Young}\ \emph {et~al.}(2012)\citenamefont {Young},
  \citenamefont {Zaheer}, \citenamefont {Teo}, \citenamefont {Kane},
  \citenamefont {Mele},\ and\ \citenamefont {Rappe}}]{young2012dirac}%
  \BibitemOpen
  \bibfield  {author} {\bibinfo {author} {\bibfnamefont {S.~M.}\ \bibnamefont
  {Young}}, \bibinfo {author} {\bibfnamefont {S.}~\bibnamefont {Zaheer}},
  \bibinfo {author} {\bibfnamefont {J.~C.}\ \bibnamefont {Teo}}, \bibinfo
  {author} {\bibfnamefont {C.~L.}\ \bibnamefont {Kane}}, \bibinfo {author}
  {\bibfnamefont {E.~J.}\ \bibnamefont {Mele}},\ and\ \bibinfo {author}
  {\bibfnamefont {A.~M.}\ \bibnamefont {Rappe}},\ }\href@noop {} {\bibfield
  {journal} {\bibinfo  {journal} {Physical review letters}\ }\textbf {\bibinfo
  {volume} {108}},\ \bibinfo {pages} {140405} (\bibinfo {year}
  {2012})}\BibitemShut {NoStop}%
\bibitem [{\citenamefont {Wang}\ \emph {et~al.}(2012)\citenamefont {Wang},
  \citenamefont {Sun}, \citenamefont {Chen}, \citenamefont {Franchini},
  \citenamefont {Xu}, \citenamefont {Weng}, \citenamefont {Dai},\ and\
  \citenamefont {Fang}}]{wang2012dirac}%
  \BibitemOpen
  \bibfield  {author} {\bibinfo {author} {\bibfnamefont {Z.}~\bibnamefont
  {Wang}}, \bibinfo {author} {\bibfnamefont {Y.}~\bibnamefont {Sun}}, \bibinfo
  {author} {\bibfnamefont {X.-Q.}\ \bibnamefont {Chen}}, \bibinfo {author}
  {\bibfnamefont {C.}~\bibnamefont {Franchini}}, \bibinfo {author}
  {\bibfnamefont {G.}~\bibnamefont {Xu}}, \bibinfo {author} {\bibfnamefont
  {H.}~\bibnamefont {Weng}}, \bibinfo {author} {\bibfnamefont {X.}~\bibnamefont
  {Dai}},\ and\ \bibinfo {author} {\bibfnamefont {Z.}~\bibnamefont {Fang}},\
  }\href@noop {} {\bibfield  {journal} {\bibinfo  {journal} {Physical Review
  B}\ }\textbf {\bibinfo {volume} {85}},\ \bibinfo {pages} {195320} (\bibinfo
  {year} {2012})}\BibitemShut {NoStop}%
\bibitem [{\citenamefont {Wang}\ \emph {et~al.}(2013)\citenamefont {Wang},
  \citenamefont {Weng}, \citenamefont {Wu}, \citenamefont {Dai},\ and\
  \citenamefont {Fang}}]{wang2013three}%
  \BibitemOpen
  \bibfield  {author} {\bibinfo {author} {\bibfnamefont {Z.}~\bibnamefont
  {Wang}}, \bibinfo {author} {\bibfnamefont {H.}~\bibnamefont {Weng}}, \bibinfo
  {author} {\bibfnamefont {Q.}~\bibnamefont {Wu}}, \bibinfo {author}
  {\bibfnamefont {X.}~\bibnamefont {Dai}},\ and\ \bibinfo {author}
  {\bibfnamefont {Z.}~\bibnamefont {Fang}},\ }\href@noop {} {\bibfield
  {journal} {\bibinfo  {journal} {Physical Review B}\ }\textbf {\bibinfo
  {volume} {88}},\ \bibinfo {pages} {125427} (\bibinfo {year}
  {2013})}\BibitemShut {NoStop}%
\bibitem [{\citenamefont {Yang}\ and\ \citenamefont
  {Nagaosa}(2014)}]{yang2014classification}%
  \BibitemOpen
  \bibfield  {author} {\bibinfo {author} {\bibfnamefont {B.-J.}\ \bibnamefont
  {Yang}}\ and\ \bibinfo {author} {\bibfnamefont {N.}~\bibnamefont {Nagaosa}},\
  }\href@noop {} {\bibfield  {journal} {\bibinfo  {journal} {Nature
  communications}\ }\textbf {\bibinfo {volume} {5}},\ \bibinfo {pages} {4898}
  (\bibinfo {year} {2014})}\BibitemShut {NoStop}%
\bibitem [{\citenamefont {Li}\ \emph {et~al.}(2022)\citenamefont {Li},
  \citenamefont {Cheung},\ and\ \citenamefont {Xu}}]{li2022topological}%
  \BibitemOpen
  \bibfield  {author} {\bibinfo {author} {\bibfnamefont {Y.}~\bibnamefont
  {Li}}, \bibinfo {author} {\bibfnamefont {C.-H.}\ \bibnamefont {Cheung}},\
  and\ \bibinfo {author} {\bibfnamefont {G.}~\bibnamefont {Xu}},\ }\href@noop
  {} {\bibfield  {journal} {\bibinfo  {journal} {Physical Review B}\ }\textbf
  {\bibinfo {volume} {105}},\ \bibinfo {pages} {035136} (\bibinfo {year}
  {2022})}\BibitemShut {NoStop}%
\bibitem [{\citenamefont {Burkov}(2016)}]{burkov2016topological}%
  \BibitemOpen
  \bibfield  {author} {\bibinfo {author} {\bibfnamefont {A.}~\bibnamefont
  {Burkov}},\ }\href@noop {} {\bibfield  {journal} {\bibinfo  {journal} {Nature
  materials}\ }\textbf {\bibinfo {volume} {15}},\ \bibinfo {pages} {1145}
  (\bibinfo {year} {2016})}\BibitemShut {NoStop}%
\bibitem [{\citenamefont {Weng}\ \emph {et~al.}(2016)\citenamefont {Weng},
  \citenamefont {Fang}, \citenamefont {Fang},\ and\ \citenamefont
  {Dai}}]{weng2016topological}%
  \BibitemOpen
  \bibfield  {author} {\bibinfo {author} {\bibfnamefont {H.}~\bibnamefont
  {Weng}}, \bibinfo {author} {\bibfnamefont {C.}~\bibnamefont {Fang}}, \bibinfo
  {author} {\bibfnamefont {Z.}~\bibnamefont {Fang}},\ and\ \bibinfo {author}
  {\bibfnamefont {X.}~\bibnamefont {Dai}},\ }\href@noop {} {\bibfield
  {journal} {\bibinfo  {journal} {Physical Review B}\ }\textbf {\bibinfo
  {volume} {93}},\ \bibinfo {pages} {241202} (\bibinfo {year}
  {2016})}\BibitemShut {NoStop}%
\bibitem [{\citenamefont {Bradlyn}\ \emph {et~al.}(2016)\citenamefont
  {Bradlyn}, \citenamefont {Cano}, \citenamefont {Wang}, \citenamefont
  {Vergniory}, \citenamefont {Felser}, \citenamefont {Cava},\ and\
  \citenamefont {Bernevig}}]{bradlyn2016beyond}%
  \BibitemOpen
  \bibfield  {author} {\bibinfo {author} {\bibfnamefont {B.}~\bibnamefont
  {Bradlyn}}, \bibinfo {author} {\bibfnamefont {J.}~\bibnamefont {Cano}},
  \bibinfo {author} {\bibfnamefont {Z.}~\bibnamefont {Wang}}, \bibinfo {author}
  {\bibfnamefont {M.}~\bibnamefont {Vergniory}}, \bibinfo {author}
  {\bibfnamefont {C.}~\bibnamefont {Felser}}, \bibinfo {author} {\bibfnamefont
  {R.~J.}\ \bibnamefont {Cava}},\ and\ \bibinfo {author} {\bibfnamefont
  {B.~A.}\ \bibnamefont {Bernevig}},\ }\href@noop {} {\bibfield  {journal}
  {\bibinfo  {journal} {Science}\ }\textbf {\bibinfo {volume} {353}},\ \bibinfo
  {pages} {aaf5037} (\bibinfo {year} {2016})}\BibitemShut {NoStop}%
\bibitem [{\citenamefont {Tang}\ \emph {et~al.}(2017)\citenamefont {Tang},
  \citenamefont {Zhou},\ and\ \citenamefont {Zhang}}]{tang2017multiple}%
  \BibitemOpen
  \bibfield  {author} {\bibinfo {author} {\bibfnamefont {P.}~\bibnamefont
  {Tang}}, \bibinfo {author} {\bibfnamefont {Q.}~\bibnamefont {Zhou}},\ and\
  \bibinfo {author} {\bibfnamefont {S.-C.}\ \bibnamefont {Zhang}},\ }\href@noop
  {} {\bibfield  {journal} {\bibinfo  {journal} {Physical review letters}\
  }\textbf {\bibinfo {volume} {119}},\ \bibinfo {pages} {206402} (\bibinfo
  {year} {2017})}\BibitemShut {NoStop}%
\bibitem [{\citenamefont {Zhu}\ \emph {et~al.}(2016)\citenamefont {Zhu},
  \citenamefont {Winkler}, \citenamefont {Wu}, \citenamefont {Li},\ and\
  \citenamefont {Soluyanov}}]{zhu2016triple}%
  \BibitemOpen
  \bibfield  {author} {\bibinfo {author} {\bibfnamefont {Z.}~\bibnamefont
  {Zhu}}, \bibinfo {author} {\bibfnamefont {G.~W.}\ \bibnamefont {Winkler}},
  \bibinfo {author} {\bibfnamefont {Q.}~\bibnamefont {Wu}}, \bibinfo {author}
  {\bibfnamefont {J.}~\bibnamefont {Li}},\ and\ \bibinfo {author}
  {\bibfnamefont {A.~A.}\ \bibnamefont {Soluyanov}},\ }\href@noop {} {\bibfield
   {journal} {\bibinfo  {journal} {Physical Review X}\ }\textbf {\bibinfo
  {volume} {6}},\ \bibinfo {pages} {031003} (\bibinfo {year}
  {2016})}\BibitemShut {NoStop}%
\bibitem [{\citenamefont {Wang}\ \emph {et~al.}(2017)\citenamefont {Wang},
  \citenamefont {Sui}, \citenamefont {Shi}, \citenamefont {Pan}, \citenamefont
  {Zhang}, \citenamefont {Liu}, \citenamefont {Wei}, \citenamefont {Yan},\ and\
  \citenamefont {Huang}}]{wang2017prediction}%
  \BibitemOpen
  \bibfield  {author} {\bibinfo {author} {\bibfnamefont {J.}~\bibnamefont
  {Wang}}, \bibinfo {author} {\bibfnamefont {X.}~\bibnamefont {Sui}}, \bibinfo
  {author} {\bibfnamefont {W.}~\bibnamefont {Shi}}, \bibinfo {author}
  {\bibfnamefont {J.}~\bibnamefont {Pan}}, \bibinfo {author} {\bibfnamefont
  {S.}~\bibnamefont {Zhang}}, \bibinfo {author} {\bibfnamefont
  {F.}~\bibnamefont {Liu}}, \bibinfo {author} {\bibfnamefont {S.-H.}\
  \bibnamefont {Wei}}, \bibinfo {author} {\bibfnamefont {Q.}~\bibnamefont
  {Yan}},\ and\ \bibinfo {author} {\bibfnamefont {B.}~\bibnamefont {Huang}},\
  }\href@noop {} {\bibfield  {journal} {\bibinfo  {journal} {Physical Review
  Letters}\ }\textbf {\bibinfo {volume} {119}},\ \bibinfo {pages} {256402}
  (\bibinfo {year} {2017})}\BibitemShut {NoStop}%
\bibitem [{\citenamefont {Mondal}\ \emph {et~al.}(2019)\citenamefont {Mondal},
  \citenamefont {Barman}, \citenamefont {Alam},\ and\ \citenamefont
  {Pathak}}]{mondal2019broken}%
  \BibitemOpen
  \bibfield  {author} {\bibinfo {author} {\bibfnamefont {C.}~\bibnamefont
  {Mondal}}, \bibinfo {author} {\bibfnamefont {C.}~\bibnamefont {Barman}},
  \bibinfo {author} {\bibfnamefont {A.}~\bibnamefont {Alam}},\ and\ \bibinfo
  {author} {\bibfnamefont {B.}~\bibnamefont {Pathak}},\ }\href@noop {}
  {\bibfield  {journal} {\bibinfo  {journal} {Physical Review B}\ }\textbf
  {\bibinfo {volume} {99}},\ \bibinfo {pages} {205112} (\bibinfo {year}
  {2019})}\BibitemShut {NoStop}%
\bibitem [{\citenamefont {B{\"u}ttiker}(1988)}]{buttiker1988absence}%
  \BibitemOpen
  \bibfield  {author} {\bibinfo {author} {\bibfnamefont {M.}~\bibnamefont
  {B{\"u}ttiker}},\ }\href@noop {} {\bibfield  {journal} {\bibinfo  {journal}
  {Physical Review B}\ }\textbf {\bibinfo {volume} {38}},\ \bibinfo {pages}
  {9375} (\bibinfo {year} {1988})}\BibitemShut {NoStop}%
\bibitem [{\citenamefont {Hsieh}\ \emph {et~al.}(2009)\citenamefont {Hsieh},
  \citenamefont {Xia}, \citenamefont {Qian}, \citenamefont {Wray},
  \citenamefont {Dil}, \citenamefont {Meier}, \citenamefont {Osterwalder},
  \citenamefont {Patthey}, \citenamefont {Checkelsky}, \citenamefont {Ong}
  \emph {et~al.}}]{hsieh2009tunable}%
  \BibitemOpen
  \bibfield  {author} {\bibinfo {author} {\bibfnamefont {D.}~\bibnamefont
  {Hsieh}}, \bibinfo {author} {\bibfnamefont {Y.}~\bibnamefont {Xia}}, \bibinfo
  {author} {\bibfnamefont {D.}~\bibnamefont {Qian}}, \bibinfo {author}
  {\bibfnamefont {L.}~\bibnamefont {Wray}}, \bibinfo {author} {\bibfnamefont
  {J.}~\bibnamefont {Dil}}, \bibinfo {author} {\bibfnamefont {F.}~\bibnamefont
  {Meier}}, \bibinfo {author} {\bibfnamefont {J.}~\bibnamefont {Osterwalder}},
  \bibinfo {author} {\bibfnamefont {L.}~\bibnamefont {Patthey}}, \bibinfo
  {author} {\bibfnamefont {J.}~\bibnamefont {Checkelsky}}, \bibinfo {author}
  {\bibfnamefont {N.~P.}\ \bibnamefont {Ong}}, \emph {et~al.},\ }\href@noop {}
  {\bibfield  {journal} {\bibinfo  {journal} {Nature}\ }\textbf {\bibinfo
  {volume} {460}},\ \bibinfo {pages} {1101} (\bibinfo {year}
  {2009})}\BibitemShut {NoStop}%
\bibitem [{\citenamefont {Weng}\ \emph {et~al.}(2015)\citenamefont {Weng},
  \citenamefont {Fang}, \citenamefont {Fang}, \citenamefont {Bernevig},\ and\
  \citenamefont {Dai}}]{weng2015weyl}%
  \BibitemOpen
  \bibfield  {author} {\bibinfo {author} {\bibfnamefont {H.}~\bibnamefont
  {Weng}}, \bibinfo {author} {\bibfnamefont {C.}~\bibnamefont {Fang}}, \bibinfo
  {author} {\bibfnamefont {Z.}~\bibnamefont {Fang}}, \bibinfo {author}
  {\bibfnamefont {B.~A.}\ \bibnamefont {Bernevig}},\ and\ \bibinfo {author}
  {\bibfnamefont {X.}~\bibnamefont {Dai}},\ }\href@noop {} {\bibfield
  {journal} {\bibinfo  {journal} {Physical Review X}\ }\textbf {\bibinfo
  {volume} {5}},\ \bibinfo {pages} {011029} (\bibinfo {year}
  {2015})}\BibitemShut {NoStop}%
\bibitem [{\citenamefont {Lv}\ \emph {et~al.}(2015)\citenamefont {Lv},
  \citenamefont {Weng}, \citenamefont {Fu}, \citenamefont {Wang}, \citenamefont
  {Miao}, \citenamefont {Ma}, \citenamefont {Richard}, \citenamefont {Huang},
  \citenamefont {Zhao}, \citenamefont {Chen} \emph
  {et~al.}}]{lv2015experimental}%
  \BibitemOpen
  \bibfield  {author} {\bibinfo {author} {\bibfnamefont {B.}~\bibnamefont
  {Lv}}, \bibinfo {author} {\bibfnamefont {H.}~\bibnamefont {Weng}}, \bibinfo
  {author} {\bibfnamefont {B.}~\bibnamefont {Fu}}, \bibinfo {author}
  {\bibfnamefont {X.~P.}\ \bibnamefont {Wang}}, \bibinfo {author}
  {\bibfnamefont {H.}~\bibnamefont {Miao}}, \bibinfo {author} {\bibfnamefont
  {J.}~\bibnamefont {Ma}}, \bibinfo {author} {\bibfnamefont {P.}~\bibnamefont
  {Richard}}, \bibinfo {author} {\bibfnamefont {X.}~\bibnamefont {Huang}},
  \bibinfo {author} {\bibfnamefont {L.}~\bibnamefont {Zhao}}, \bibinfo {author}
  {\bibfnamefont {G.}~\bibnamefont {Chen}}, \emph {et~al.},\ }\href@noop {}
  {\bibfield  {journal} {\bibinfo  {journal} {Physical Review X}\ }\textbf
  {\bibinfo {volume} {5}},\ \bibinfo {pages} {031013} (\bibinfo {year}
  {2015})}\BibitemShut {NoStop}%
\bibitem [{\citenamefont {Li}\ \emph {et~al.}(2016)\citenamefont {Li},
  \citenamefont {He}, \citenamefont {Lu}, \citenamefont {Zhang}, \citenamefont
  {Liu}, \citenamefont {Ma}, \citenamefont {Fan}, \citenamefont {Shen},\ and\
  \citenamefont {Wang}}]{li2016negative}%
  \BibitemOpen
  \bibfield  {author} {\bibinfo {author} {\bibfnamefont {H.}~\bibnamefont
  {Li}}, \bibinfo {author} {\bibfnamefont {H.}~\bibnamefont {He}}, \bibinfo
  {author} {\bibfnamefont {H.-Z.}\ \bibnamefont {Lu}}, \bibinfo {author}
  {\bibfnamefont {H.}~\bibnamefont {Zhang}}, \bibinfo {author} {\bibfnamefont
  {H.}~\bibnamefont {Liu}}, \bibinfo {author} {\bibfnamefont {R.}~\bibnamefont
  {Ma}}, \bibinfo {author} {\bibfnamefont {Z.}~\bibnamefont {Fan}}, \bibinfo
  {author} {\bibfnamefont {S.-Q.}\ \bibnamefont {Shen}},\ and\ \bibinfo
  {author} {\bibfnamefont {J.}~\bibnamefont {Wang}},\ }\href@noop {} {\bibfield
   {journal} {\bibinfo  {journal} {Nature communications}\ }\textbf {\bibinfo
  {volume} {7}},\ \bibinfo {pages} {10301} (\bibinfo {year}
  {2016})}\BibitemShut {NoStop}%
\bibitem [{\citenamefont {Mellnik}\ \emph {et~al.}(2014)\citenamefont
  {Mellnik}, \citenamefont {Lee}, \citenamefont {Richardella}, \citenamefont
  {Grab}, \citenamefont {Mintun}, \citenamefont {Fischer}, \citenamefont
  {Vaezi}, \citenamefont {Manchon}, \citenamefont {Kim}, \citenamefont
  {Samarth} \emph {et~al.}}]{mellnik2014spin}%
  \BibitemOpen
  \bibfield  {author} {\bibinfo {author} {\bibfnamefont {A.}~\bibnamefont
  {Mellnik}}, \bibinfo {author} {\bibfnamefont {J.}~\bibnamefont {Lee}},
  \bibinfo {author} {\bibfnamefont {A.}~\bibnamefont {Richardella}}, \bibinfo
  {author} {\bibfnamefont {J.}~\bibnamefont {Grab}}, \bibinfo {author}
  {\bibfnamefont {P.}~\bibnamefont {Mintun}}, \bibinfo {author} {\bibfnamefont
  {M.~H.}\ \bibnamefont {Fischer}}, \bibinfo {author} {\bibfnamefont
  {A.}~\bibnamefont {Vaezi}}, \bibinfo {author} {\bibfnamefont
  {A.}~\bibnamefont {Manchon}}, \bibinfo {author} {\bibfnamefont {E.-A.}\
  \bibnamefont {Kim}}, \bibinfo {author} {\bibfnamefont {N.}~\bibnamefont
  {Samarth}}, \emph {et~al.},\ }\href@noop {} {\bibfield  {journal} {\bibinfo
  {journal} {Nature}\ }\textbf {\bibinfo {volume} {511}},\ \bibinfo {pages}
  {449} (\bibinfo {year} {2014})}\BibitemShut {NoStop}%
\bibitem [{\citenamefont {Tang}\ \emph {et~al.}(2016)\citenamefont {Tang},
  \citenamefont {Zhou}, \citenamefont {Xu},\ and\ \citenamefont
  {Zhang}}]{tang2016dirac}%
  \BibitemOpen
  \bibfield  {author} {\bibinfo {author} {\bibfnamefont {P.}~\bibnamefont
  {Tang}}, \bibinfo {author} {\bibfnamefont {Q.}~\bibnamefont {Zhou}}, \bibinfo
  {author} {\bibfnamefont {G.}~\bibnamefont {Xu}},\ and\ \bibinfo {author}
  {\bibfnamefont {S.-C.}\ \bibnamefont {Zhang}},\ }\href@noop {} {\bibfield
  {journal} {\bibinfo  {journal} {Nature Physics}\ }\textbf {\bibinfo {volume}
  {12}},\ \bibinfo {pages} {1100} (\bibinfo {year} {2016})}\BibitemShut
  {NoStop}%
\bibitem [{\citenamefont {Hua}\ \emph {et~al.}(2018)\citenamefont {Hua},
  \citenamefont {Nie}, \citenamefont {Song}, \citenamefont {Yu}, \citenamefont
  {Xu},\ and\ \citenamefont {Yao}}]{hua2018dirac}%
  \BibitemOpen
  \bibfield  {author} {\bibinfo {author} {\bibfnamefont {G.}~\bibnamefont
  {Hua}}, \bibinfo {author} {\bibfnamefont {S.}~\bibnamefont {Nie}}, \bibinfo
  {author} {\bibfnamefont {Z.}~\bibnamefont {Song}}, \bibinfo {author}
  {\bibfnamefont {R.}~\bibnamefont {Yu}}, \bibinfo {author} {\bibfnamefont
  {G.}~\bibnamefont {Xu}},\ and\ \bibinfo {author} {\bibfnamefont
  {K.}~\bibnamefont {Yao}},\ }\href@noop {} {\bibfield  {journal} {\bibinfo
  {journal} {Physical Review B}\ }\textbf {\bibinfo {volume} {98}},\ \bibinfo
  {pages} {201116} (\bibinfo {year} {2018})}\BibitemShut {NoStop}%
\bibitem [{\citenamefont {Zhang}\ \emph {et~al.}(2019)\citenamefont {Zhang},
  \citenamefont {Shi}, \citenamefont {Zhu}, \citenamefont {Xing}, \citenamefont
  {Zhang},\ and\ \citenamefont {Wang}}]{zhang2019topological}%
  \BibitemOpen
  \bibfield  {author} {\bibinfo {author} {\bibfnamefont {D.}~\bibnamefont
  {Zhang}}, \bibinfo {author} {\bibfnamefont {M.}~\bibnamefont {Shi}}, \bibinfo
  {author} {\bibfnamefont {T.}~\bibnamefont {Zhu}}, \bibinfo {author}
  {\bibfnamefont {D.}~\bibnamefont {Xing}}, \bibinfo {author} {\bibfnamefont
  {H.}~\bibnamefont {Zhang}},\ and\ \bibinfo {author} {\bibfnamefont
  {J.}~\bibnamefont {Wang}},\ }\href@noop {} {\bibfield  {journal} {\bibinfo
  {journal} {Physical review letters}\ }\textbf {\bibinfo {volume} {122}},\
  \bibinfo {pages} {206401} (\bibinfo {year} {2019})}\BibitemShut {NoStop}%
\bibitem [{\citenamefont {Zou}\ \emph {et~al.}(2019)\citenamefont {Zou},
  \citenamefont {He},\ and\ \citenamefont {Xu}}]{zou2019study}%
  \BibitemOpen
  \bibfield  {author} {\bibinfo {author} {\bibfnamefont {J.}~\bibnamefont
  {Zou}}, \bibinfo {author} {\bibfnamefont {Z.}~\bibnamefont {He}},\ and\
  \bibinfo {author} {\bibfnamefont {G.}~\bibnamefont {Xu}},\ }\href@noop {}
  {\bibfield  {journal} {\bibinfo  {journal} {npj Computational Materials}\
  }\textbf {\bibinfo {volume} {5}},\ \bibinfo {pages} {96} (\bibinfo {year}
  {2019})}\BibitemShut {NoStop}%
\bibitem [{\citenamefont {Jin}\ \emph {et~al.}(2021)\citenamefont {Jin},
  \citenamefont {Zeng}, \citenamefont {Feng}, \citenamefont {Du}, \citenamefont
  {Wu}, \citenamefont {Sheng}, \citenamefont {Yu}, \citenamefont {Zhu},\ and\
  \citenamefont {Yang}}]{jin2021multiple}%
  \BibitemOpen
  \bibfield  {author} {\bibinfo {author} {\bibfnamefont {Y.}~\bibnamefont
  {Jin}}, \bibinfo {author} {\bibfnamefont {X.-T.}\ \bibnamefont {Zeng}},
  \bibinfo {author} {\bibfnamefont {X.}~\bibnamefont {Feng}}, \bibinfo {author}
  {\bibfnamefont {X.}~\bibnamefont {Du}}, \bibinfo {author} {\bibfnamefont
  {W.}~\bibnamefont {Wu}}, \bibinfo {author} {\bibfnamefont {X.-L.}\
  \bibnamefont {Sheng}}, \bibinfo {author} {\bibfnamefont {Z.-M.}\ \bibnamefont
  {Yu}}, \bibinfo {author} {\bibfnamefont {Z.}~\bibnamefont {Zhu}},\ and\
  \bibinfo {author} {\bibfnamefont {S.~A.}\ \bibnamefont {Yang}},\ }\href@noop
  {} {\bibfield  {journal} {\bibinfo  {journal} {Physical Review B}\ }\textbf
  {\bibinfo {volume} {104}},\ \bibinfo {pages} {165424} (\bibinfo {year}
  {2021})}\BibitemShut {NoStop}%
\bibitem [{\citenamefont {Malick}\ \emph {et~al.}(2022)\citenamefont {Malick},
  \citenamefont {Singh}, \citenamefont {Laha}, \citenamefont {Kanchana},
  \citenamefont {Hossain},\ and\ \citenamefont
  {Kaczorowski}}]{malick2022electronic}%
  \BibitemOpen
  \bibfield  {author} {\bibinfo {author} {\bibfnamefont {S.}~\bibnamefont
  {Malick}}, \bibinfo {author} {\bibfnamefont {J.}~\bibnamefont {Singh}},
  \bibinfo {author} {\bibfnamefont {A.}~\bibnamefont {Laha}}, \bibinfo {author}
  {\bibfnamefont {V.}~\bibnamefont {Kanchana}}, \bibinfo {author}
  {\bibfnamefont {Z.}~\bibnamefont {Hossain}},\ and\ \bibinfo {author}
  {\bibfnamefont {D.}~\bibnamefont {Kaczorowski}},\ }\href@noop {} {\bibfield
  {journal} {\bibinfo  {journal} {Physical Review B}\ }\textbf {\bibinfo
  {volume} {105}},\ \bibinfo {pages} {045103} (\bibinfo {year}
  {2022})}\BibitemShut {NoStop}%
\bibitem [{\citenamefont {Chang}\ \emph {et~al.}(2023)\citenamefont {Chang},
  \citenamefont {Liu},\ and\ \citenamefont {MacDonald}}]{chang2023colloquium}%
  \BibitemOpen
  \bibfield  {author} {\bibinfo {author} {\bibfnamefont {C.-Z.}\ \bibnamefont
  {Chang}}, \bibinfo {author} {\bibfnamefont {C.-X.}\ \bibnamefont {Liu}},\
  and\ \bibinfo {author} {\bibfnamefont {A.~H.}\ \bibnamefont {MacDonald}},\
  }\href@noop {} {\bibfield  {journal} {\bibinfo  {journal} {Reviews of Modern
  Physics}\ }\textbf {\bibinfo {volume} {95}},\ \bibinfo {pages} {011002}
  (\bibinfo {year} {2023})}\BibitemShut {NoStop}%
\bibitem [{\citenamefont {Xu}\ \emph {et~al.}(2019)\citenamefont {Xu},
  \citenamefont {Song}, \citenamefont {Wang}, \citenamefont {Weng},\ and\
  \citenamefont {Dai}}]{xu2019higher}%
  \BibitemOpen
  \bibfield  {author} {\bibinfo {author} {\bibfnamefont {Y.}~\bibnamefont
  {Xu}}, \bibinfo {author} {\bibfnamefont {Z.}~\bibnamefont {Song}}, \bibinfo
  {author} {\bibfnamefont {Z.}~\bibnamefont {Wang}}, \bibinfo {author}
  {\bibfnamefont {H.}~\bibnamefont {Weng}},\ and\ \bibinfo {author}
  {\bibfnamefont {X.}~\bibnamefont {Dai}},\ }\href@noop {} {\bibfield
  {journal} {\bibinfo  {journal} {Physical review letters}\ }\textbf {\bibinfo
  {volume} {122}},\ \bibinfo {pages} {256402} (\bibinfo {year}
  {2019})}\BibitemShut {NoStop}%
\bibitem [{\citenamefont {Deng}\ \emph {et~al.}(2020)\citenamefont {Deng},
  \citenamefont {Yu}, \citenamefont {Shi}, \citenamefont {Guo}, \citenamefont
  {Xu}, \citenamefont {Wang}, \citenamefont {Chen},\ and\ \citenamefont
  {Zhang}}]{deng2020quantum}%
  \BibitemOpen
  \bibfield  {author} {\bibinfo {author} {\bibfnamefont {Y.}~\bibnamefont
  {Deng}}, \bibinfo {author} {\bibfnamefont {Y.}~\bibnamefont {Yu}}, \bibinfo
  {author} {\bibfnamefont {M.~Z.}\ \bibnamefont {Shi}}, \bibinfo {author}
  {\bibfnamefont {Z.}~\bibnamefont {Guo}}, \bibinfo {author} {\bibfnamefont
  {Z.}~\bibnamefont {Xu}}, \bibinfo {author} {\bibfnamefont {J.}~\bibnamefont
  {Wang}}, \bibinfo {author} {\bibfnamefont {X.~H.}\ \bibnamefont {Chen}},\
  and\ \bibinfo {author} {\bibfnamefont {Y.}~\bibnamefont {Zhang}},\
  }\href@noop {} {\bibfield  {journal} {\bibinfo  {journal} {Science}\ }\textbf
  {\bibinfo {volume} {367}},\ \bibinfo {pages} {895} (\bibinfo {year}
  {2020})}\BibitemShut {NoStop}%
\bibitem [{\citenamefont {Liu}\ \emph {et~al.}(2019)\citenamefont {Liu},
  \citenamefont {Liang}, \citenamefont {Liu}, \citenamefont {Xu}, \citenamefont
  {Li}, \citenamefont {Chen}, \citenamefont {Pei}, \citenamefont {Shi},
  \citenamefont {Mo}, \citenamefont {Dudin} \emph {et~al.}}]{liu2019magnetic}%
  \BibitemOpen
  \bibfield  {author} {\bibinfo {author} {\bibfnamefont {D.}~\bibnamefont
  {Liu}}, \bibinfo {author} {\bibfnamefont {A.}~\bibnamefont {Liang}}, \bibinfo
  {author} {\bibfnamefont {E.}~\bibnamefont {Liu}}, \bibinfo {author}
  {\bibfnamefont {Q.}~\bibnamefont {Xu}}, \bibinfo {author} {\bibfnamefont
  {Y.}~\bibnamefont {Li}}, \bibinfo {author} {\bibfnamefont {C.}~\bibnamefont
  {Chen}}, \bibinfo {author} {\bibfnamefont {D.}~\bibnamefont {Pei}}, \bibinfo
  {author} {\bibfnamefont {W.}~\bibnamefont {Shi}}, \bibinfo {author}
  {\bibfnamefont {S.}~\bibnamefont {Mo}}, \bibinfo {author} {\bibfnamefont
  {P.}~\bibnamefont {Dudin}}, \emph {et~al.},\ }\href@noop {} {\bibfield
  {journal} {\bibinfo  {journal} {Science}\ }\textbf {\bibinfo {volume}
  {365}},\ \bibinfo {pages} {1282} (\bibinfo {year} {2019})}\BibitemShut
  {NoStop}%
\bibitem [{\citenamefont {Ma}\ \emph {et~al.}(2019)\citenamefont {Ma},
  \citenamefont {Nie}, \citenamefont {Yi}, \citenamefont {Jandke},
  \citenamefont {Shang}, \citenamefont {Yao}, \citenamefont {Naamneh},
  \citenamefont {Yan}, \citenamefont {Sun}, \citenamefont {Chikina} \emph
  {et~al.}}]{ma2019spin}%
  \BibitemOpen
  \bibfield  {author} {\bibinfo {author} {\bibfnamefont {J.-Z.}\ \bibnamefont
  {Ma}}, \bibinfo {author} {\bibfnamefont {S.}~\bibnamefont {Nie}}, \bibinfo
  {author} {\bibfnamefont {C.}~\bibnamefont {Yi}}, \bibinfo {author}
  {\bibfnamefont {J.}~\bibnamefont {Jandke}}, \bibinfo {author} {\bibfnamefont
  {T.}~\bibnamefont {Shang}}, \bibinfo {author} {\bibfnamefont {M.-Y.}\
  \bibnamefont {Yao}}, \bibinfo {author} {\bibfnamefont {M.}~\bibnamefont
  {Naamneh}}, \bibinfo {author} {\bibfnamefont {L.}~\bibnamefont {Yan}},
  \bibinfo {author} {\bibfnamefont {Y.}~\bibnamefont {Sun}}, \bibinfo {author}
  {\bibfnamefont {A.}~\bibnamefont {Chikina}}, \emph {et~al.},\ }\href@noop {}
  {\bibfield  {journal} {\bibinfo  {journal} {Science advances}\ }\textbf
  {\bibinfo {volume} {5}},\ \bibinfo {pages} {eaaw4718} (\bibinfo {year}
  {2019})}\BibitemShut {NoStop}%
\bibitem [{\citenamefont {{\v{S}}mejkal}\ \emph {et~al.}(2018)\citenamefont
  {{\v{S}}mejkal}, \citenamefont {Mokrousov}, \citenamefont {Yan},\ and\
  \citenamefont {MacDonald}}]{vsmejkal2018topological}%
  \BibitemOpen
  \bibfield  {author} {\bibinfo {author} {\bibfnamefont {L.}~\bibnamefont
  {{\v{S}}mejkal}}, \bibinfo {author} {\bibfnamefont {Y.}~\bibnamefont
  {Mokrousov}}, \bibinfo {author} {\bibfnamefont {B.}~\bibnamefont {Yan}},\
  and\ \bibinfo {author} {\bibfnamefont {A.~H.}\ \bibnamefont {MacDonald}},\
  }\href@noop {} {\bibfield  {journal} {\bibinfo  {journal} {Nature physics}\
  }\textbf {\bibinfo {volume} {14}},\ \bibinfo {pages} {242} (\bibinfo {year}
  {2018})}\BibitemShut {NoStop}%
\bibitem [{\citenamefont {He}\ \emph {et~al.}(2022)\citenamefont {He},
  \citenamefont {Hughes}, \citenamefont {Armitage}, \citenamefont {Tokura},\
  and\ \citenamefont {Wang}}]{he2022topological}%
  \BibitemOpen
  \bibfield  {author} {\bibinfo {author} {\bibfnamefont {Q.~L.}\ \bibnamefont
  {He}}, \bibinfo {author} {\bibfnamefont {T.~L.}\ \bibnamefont {Hughes}},
  \bibinfo {author} {\bibfnamefont {N.~P.}\ \bibnamefont {Armitage}}, \bibinfo
  {author} {\bibfnamefont {Y.}~\bibnamefont {Tokura}},\ and\ \bibinfo {author}
  {\bibfnamefont {K.~L.}\ \bibnamefont {Wang}},\ }\href@noop {} {\bibfield
  {journal} {\bibinfo  {journal} {Nature materials}\ }\textbf {\bibinfo
  {volume} {21}},\ \bibinfo {pages} {15} (\bibinfo {year} {2022})}\BibitemShut
  {NoStop}%
\bibitem [{\citenamefont {Bernevig}\ \emph {et~al.}(2022)\citenamefont
  {Bernevig}, \citenamefont {Felser},\ and\ \citenamefont
  {Beidenkopf}}]{bernevig2022progress}%
  \BibitemOpen
  \bibfield  {author} {\bibinfo {author} {\bibfnamefont {B.~A.}\ \bibnamefont
  {Bernevig}}, \bibinfo {author} {\bibfnamefont {C.}~\bibnamefont {Felser}},\
  and\ \bibinfo {author} {\bibfnamefont {H.}~\bibnamefont {Beidenkopf}},\
  }\href@noop {} {\bibfield  {journal} {\bibinfo  {journal} {Nature}\ }\textbf
  {\bibinfo {volume} {603}},\ \bibinfo {pages} {41} (\bibinfo {year}
  {2022})}\BibitemShut {NoStop}%
\bibitem [{\citenamefont {Takahashi}\ \emph {et~al.}(2023)\citenamefont
  {Takahashi}, \citenamefont {Akiba}, \citenamefont {Takahashi}, \citenamefont
  {Mayo}, \citenamefont {Ochi}, \citenamefont {Kobayashi},\ and\ \citenamefont
  {Ishiwata}}]{takahashi2023superconductivity}%
  \BibitemOpen
  \bibfield  {author} {\bibinfo {author} {\bibfnamefont {H.}~\bibnamefont
  {Takahashi}}, \bibinfo {author} {\bibfnamefont {K.}~\bibnamefont {Akiba}},
  \bibinfo {author} {\bibfnamefont {M.}~\bibnamefont {Takahashi}}, \bibinfo
  {author} {\bibfnamefont {A.~H.}\ \bibnamefont {Mayo}}, \bibinfo {author}
  {\bibfnamefont {M.}~\bibnamefont {Ochi}}, \bibinfo {author} {\bibfnamefont
  {T.~C.}\ \bibnamefont {Kobayashi}},\ and\ \bibinfo {author} {\bibfnamefont
  {S.}~\bibnamefont {Ishiwata}},\ }\href@noop {} {\bibfield  {journal}
  {\bibinfo  {journal} {journal of the physical society of japan}\ }\textbf
  {\bibinfo {volume} {92}},\ \bibinfo {pages} {013701} (\bibinfo {year}
  {2023})}\BibitemShut {NoStop}%
\bibitem [{\citenamefont {Kresse}\ and\ \citenamefont
  {Furthm{\"u}ller}(1996{\natexlab{a}})}]{kresse1996efficiency}%
  \BibitemOpen
  \bibfield  {author} {\bibinfo {author} {\bibfnamefont {G.}~\bibnamefont
  {Kresse}}\ and\ \bibinfo {author} {\bibfnamefont {J.}~\bibnamefont
  {Furthm{\"u}ller}},\ }\href@noop {} {\bibfield  {journal} {\bibinfo
  {journal} {Computational materials science}\ }\textbf {\bibinfo {volume}
  {6}},\ \bibinfo {pages} {15} (\bibinfo {year}
  {1996}{\natexlab{a}})}\BibitemShut {NoStop}%
\bibitem [{\citenamefont {Kresse}\ and\ \citenamefont
  {Furthm{\"u}ller}(1996{\natexlab{b}})}]{kresse1996efficient}%
  \BibitemOpen
  \bibfield  {author} {\bibinfo {author} {\bibfnamefont {G.}~\bibnamefont
  {Kresse}}\ and\ \bibinfo {author} {\bibfnamefont {J.}~\bibnamefont
  {Furthm{\"u}ller}},\ }\href@noop {} {\bibfield  {journal} {\bibinfo
  {journal} {Physical review B}\ }\textbf {\bibinfo {volume} {54}},\ \bibinfo
  {pages} {11169} (\bibinfo {year} {1996}{\natexlab{b}})}\BibitemShut {NoStop}%
\bibitem [{\citenamefont {Bl{\"o}chl}(1994)}]{blochl1994projector}%
  \BibitemOpen
  \bibfield  {author} {\bibinfo {author} {\bibfnamefont {P.~E.}\ \bibnamefont
  {Bl{\"o}chl}},\ }\href@noop {} {\bibfield  {journal} {\bibinfo  {journal}
  {Physical review B}\ }\textbf {\bibinfo {volume} {50}},\ \bibinfo {pages}
  {17953} (\bibinfo {year} {1994})}\BibitemShut {NoStop}%
\bibitem [{\citenamefont {Perdew}\ \emph {et~al.}(1996)\citenamefont {Perdew},
  \citenamefont {Burke},\ and\ \citenamefont
  {Ernzerhof}}]{perdew1996generalized}%
  \BibitemOpen
  \bibfield  {author} {\bibinfo {author} {\bibfnamefont {J.~P.}\ \bibnamefont
  {Perdew}}, \bibinfo {author} {\bibfnamefont {K.}~\bibnamefont {Burke}},\ and\
  \bibinfo {author} {\bibfnamefont {M.}~\bibnamefont {Ernzerhof}},\ }\href@noop
  {} {\bibfield  {journal} {\bibinfo  {journal} {Physical review letters}\
  }\textbf {\bibinfo {volume} {77}},\ \bibinfo {pages} {3865} (\bibinfo {year}
  {1996})}\BibitemShut {NoStop}%
\bibitem [{\citenamefont {Heyd}\ \emph {et~al.}(2003)\citenamefont {Heyd},
  \citenamefont {Scuseria},\ and\ \citenamefont {Ernzerhof}}]{heyd2003hybrid}%
  \BibitemOpen
  \bibfield  {author} {\bibinfo {author} {\bibfnamefont {J.}~\bibnamefont
  {Heyd}}, \bibinfo {author} {\bibfnamefont {G.~E.}\ \bibnamefont {Scuseria}},\
  and\ \bibinfo {author} {\bibfnamefont {M.}~\bibnamefont {Ernzerhof}},\
  }\href@noop {} {\bibfield  {journal} {\bibinfo  {journal} {The Journal of
  chemical physics}\ }\textbf {\bibinfo {volume} {118}},\ \bibinfo {pages}
  {8207} (\bibinfo {year} {2003})}\BibitemShut {NoStop}%
\bibitem [{\citenamefont {Mostofi}\ \emph {et~al.}(2014)\citenamefont
  {Mostofi}, \citenamefont {Yates}, \citenamefont {Pizzi}, \citenamefont {Lee},
  \citenamefont {Souza}, \citenamefont {Vanderbilt},\ and\ \citenamefont
  {Marzari}}]{mostofi2014updated}%
  \BibitemOpen
  \bibfield  {author} {\bibinfo {author} {\bibfnamefont {A.~A.}\ \bibnamefont
  {Mostofi}}, \bibinfo {author} {\bibfnamefont {J.~R.}\ \bibnamefont {Yates}},
  \bibinfo {author} {\bibfnamefont {G.}~\bibnamefont {Pizzi}}, \bibinfo
  {author} {\bibfnamefont {Y.-S.}\ \bibnamefont {Lee}}, \bibinfo {author}
  {\bibfnamefont {I.}~\bibnamefont {Souza}}, \bibinfo {author} {\bibfnamefont
  {D.}~\bibnamefont {Vanderbilt}},\ and\ \bibinfo {author} {\bibfnamefont
  {N.}~\bibnamefont {Marzari}},\ }\href@noop {} {\bibfield  {journal} {\bibinfo
   {journal} {Computer Physics Communications}\ }\textbf {\bibinfo {volume}
  {185}},\ \bibinfo {pages} {2309} (\bibinfo {year} {2014})}\BibitemShut
  {NoStop}%
\bibitem [{\citenamefont {Zhi}\ \emph {et~al.}(2022)\citenamefont {Zhi},
  \citenamefont {Xu}, \citenamefont {Wu}, \citenamefont {Ning},\ and\
  \citenamefont {Cao}}]{zhi2022wannsymm}%
  \BibitemOpen
  \bibfield  {author} {\bibinfo {author} {\bibfnamefont {G.-X.}\ \bibnamefont
  {Zhi}}, \bibinfo {author} {\bibfnamefont {C.}~\bibnamefont {Xu}}, \bibinfo
  {author} {\bibfnamefont {S.-Q.}\ \bibnamefont {Wu}}, \bibinfo {author}
  {\bibfnamefont {F.}~\bibnamefont {Ning}},\ and\ \bibinfo {author}
  {\bibfnamefont {C.}~\bibnamefont {Cao}},\ }\href@noop {} {\bibfield
  {journal} {\bibinfo  {journal} {Computer Physics Communications}\ }\textbf
  {\bibinfo {volume} {271}},\ \bibinfo {pages} {108196} (\bibinfo {year}
  {2022})}\BibitemShut {NoStop}%
\bibitem [{\citenamefont {Wu}\ \emph {et~al.}(2018)\citenamefont {Wu},
  \citenamefont {Zhang}, \citenamefont {Song}, \citenamefont {Troyer},\ and\
  \citenamefont {Soluyanov}}]{wu2018wanniertools}%
  \BibitemOpen
  \bibfield  {author} {\bibinfo {author} {\bibfnamefont {Q.}~\bibnamefont
  {Wu}}, \bibinfo {author} {\bibfnamefont {S.}~\bibnamefont {Zhang}}, \bibinfo
  {author} {\bibfnamefont {H.-F.}\ \bibnamefont {Song}}, \bibinfo {author}
  {\bibfnamefont {M.}~\bibnamefont {Troyer}},\ and\ \bibinfo {author}
  {\bibfnamefont {A.~A.}\ \bibnamefont {Soluyanov}},\ }\href@noop {} {\bibfield
   {journal} {\bibinfo  {journal} {Computer Physics Communications}\ }\textbf
  {\bibinfo {volume} {224}},\ \bibinfo {pages} {405} (\bibinfo {year}
  {2018})}\BibitemShut {NoStop}%
\bibitem [{\citenamefont {Tomuschat}\ and\ \citenamefont
  {Schuster}(1981)}]{tomuschat1981abx}%
  \BibitemOpen
  \bibfield  {author} {\bibinfo {author} {\bibfnamefont {C.}~\bibnamefont
  {Tomuschat}}\ and\ \bibinfo {author} {\bibfnamefont {H.-U.}\ \bibnamefont
  {Schuster}},\ }\href@noop {} {\bibfield  {journal} {\bibinfo  {journal}
  {Zeitschrift f{\"u}r Naturforschung B}\ }\textbf {\bibinfo {volume} {36}},\
  \bibinfo {pages} {1193} (\bibinfo {year} {1981})}\BibitemShut {NoStop}%
\bibitem [{\citenamefont {Merlo}\ \emph {et~al.}(1990)\citenamefont {Merlo},
  \citenamefont {Pani},\ and\ \citenamefont {Fornasini}}]{merlo1990rmx}%
  \BibitemOpen
  \bibfield  {author} {\bibinfo {author} {\bibfnamefont {F.}~\bibnamefont
  {Merlo}}, \bibinfo {author} {\bibfnamefont {M.}~\bibnamefont {Pani}},\ and\
  \bibinfo {author} {\bibfnamefont {M.}~\bibnamefont {Fornasini}},\ }\href@noop
  {} {\bibfield  {journal} {\bibinfo  {journal} {Journal of the Less Common
  Metals}\ }\textbf {\bibinfo {volume} {166}},\ \bibinfo {pages} {319}
  (\bibinfo {year} {1990})}\BibitemShut {NoStop}%
\bibitem [{\citenamefont {Tomuschat}\ and\ \citenamefont
  {Schuster}(1984)}]{tomuschat1984magnetic}%
  \BibitemOpen
  \bibfield  {author} {\bibinfo {author} {\bibfnamefont {C.}~\bibnamefont
  {Tomuschat}}\ and\ \bibinfo {author} {\bibfnamefont {H.}~\bibnamefont
  {Schuster}},\ }\href@noop {} {\bibfield  {journal} {\bibinfo  {journal}
  {Zeitschrift fuer Anorganische und Allgemeine Chemie (1950)}\ }\textbf
  {\bibinfo {volume} {518}} (\bibinfo {year} {1984})}\BibitemShut {NoStop}%
\bibitem [{\citenamefont {Iha}\ \emph {et~al.}(2019)\citenamefont {Iha},
  \citenamefont {Kakihana}, \citenamefont {Matsuda}, \citenamefont {Honda},
  \citenamefont {Haga}, \citenamefont {Takeuchi}, \citenamefont {Nakashima},
  \citenamefont {Amako}, \citenamefont {Gouchi}, \citenamefont {Uwatoko} \emph
  {et~al.}}]{iha2019anomalous}%
  \BibitemOpen
  \bibfield  {author} {\bibinfo {author} {\bibfnamefont {W.}~\bibnamefont
  {Iha}}, \bibinfo {author} {\bibfnamefont {M.}~\bibnamefont {Kakihana}},
  \bibinfo {author} {\bibfnamefont {S.}~\bibnamefont {Matsuda}}, \bibinfo
  {author} {\bibfnamefont {F.}~\bibnamefont {Honda}}, \bibinfo {author}
  {\bibfnamefont {Y.}~\bibnamefont {Haga}}, \bibinfo {author} {\bibfnamefont
  {T.}~\bibnamefont {Takeuchi}}, \bibinfo {author} {\bibfnamefont
  {M.}~\bibnamefont {Nakashima}}, \bibinfo {author} {\bibfnamefont
  {Y.}~\bibnamefont {Amako}}, \bibinfo {author} {\bibfnamefont
  {J.}~\bibnamefont {Gouchi}}, \bibinfo {author} {\bibfnamefont
  {Y.}~\bibnamefont {Uwatoko}}, \emph {et~al.},\ }\href@noop {} {\bibfield
  {journal} {\bibinfo  {journal} {Journal of Alloys and Compounds}\ }\textbf
  {\bibinfo {volume} {788}},\ \bibinfo {pages} {361} (\bibinfo {year}
  {2019})}\BibitemShut {NoStop}%
\bibitem [{\citenamefont {Tong}\ \emph {et~al.}(2014)\citenamefont {Tong},
  \citenamefont {Parry}, \citenamefont {Tao}, \citenamefont {Cao},
  \citenamefont {Xu},\ and\ \citenamefont {Zeng}}]{tong2014magnetic}%
  \BibitemOpen
  \bibfield  {author} {\bibinfo {author} {\bibfnamefont {J.}~\bibnamefont
  {Tong}}, \bibinfo {author} {\bibfnamefont {J.}~\bibnamefont {Parry}},
  \bibinfo {author} {\bibfnamefont {Q.}~\bibnamefont {Tao}}, \bibinfo {author}
  {\bibfnamefont {G.-H.}\ \bibnamefont {Cao}}, \bibinfo {author} {\bibfnamefont
  {Z.-A.}\ \bibnamefont {Xu}},\ and\ \bibinfo {author} {\bibfnamefont
  {H.}~\bibnamefont {Zeng}},\ }\href@noop {} {\bibfield  {journal} {\bibinfo
  {journal} {Journal of Alloys and Compounds}\ }\textbf {\bibinfo {volume}
  {602}},\ \bibinfo {pages} {26} (\bibinfo {year} {2014})}\BibitemShut
  {NoStop}%
\bibitem [{\citenamefont {Takahashi}\ \emph {et~al.}(2020)\citenamefont
  {Takahashi}, \citenamefont {Aono}, \citenamefont {Nambu}, \citenamefont
  {Kiyanagi}, \citenamefont {Nomoto}, \citenamefont {Sakano}, \citenamefont
  {Ishizaka}, \citenamefont {Arita},\ and\ \citenamefont
  {Ishiwata}}]{takahashi2020competing}%
  \BibitemOpen
  \bibfield  {author} {\bibinfo {author} {\bibfnamefont {H.}~\bibnamefont
  {Takahashi}}, \bibinfo {author} {\bibfnamefont {K.}~\bibnamefont {Aono}},
  \bibinfo {author} {\bibfnamefont {Y.}~\bibnamefont {Nambu}}, \bibinfo
  {author} {\bibfnamefont {R.}~\bibnamefont {Kiyanagi}}, \bibinfo {author}
  {\bibfnamefont {T.}~\bibnamefont {Nomoto}}, \bibinfo {author} {\bibfnamefont
  {M.}~\bibnamefont {Sakano}}, \bibinfo {author} {\bibfnamefont
  {K.}~\bibnamefont {Ishizaka}}, \bibinfo {author} {\bibfnamefont
  {R.}~\bibnamefont {Arita}},\ and\ \bibinfo {author} {\bibfnamefont
  {S.}~\bibnamefont {Ishiwata}},\ }\href@noop {} {\bibfield  {journal}
  {\bibinfo  {journal} {Physical Review B}\ }\textbf {\bibinfo {volume}
  {102}},\ \bibinfo {pages} {174425} (\bibinfo {year} {2020})}\BibitemShut
  {NoStop}%
\bibitem [{\citenamefont {Laha}\ \emph {et~al.}(2021)\citenamefont {Laha},
  \citenamefont {Singha}, \citenamefont {Mardanya}, \citenamefont {Singh},
  \citenamefont {Agarwal}, \citenamefont {Mandal},\ and\ \citenamefont
  {Hossain}}]{laha2021topological}%
  \BibitemOpen
  \bibfield  {author} {\bibinfo {author} {\bibfnamefont {A.}~\bibnamefont
  {Laha}}, \bibinfo {author} {\bibfnamefont {R.}~\bibnamefont {Singha}},
  \bibinfo {author} {\bibfnamefont {S.}~\bibnamefont {Mardanya}}, \bibinfo
  {author} {\bibfnamefont {B.}~\bibnamefont {Singh}}, \bibinfo {author}
  {\bibfnamefont {A.}~\bibnamefont {Agarwal}}, \bibinfo {author} {\bibfnamefont
  {P.}~\bibnamefont {Mandal}},\ and\ \bibinfo {author} {\bibfnamefont
  {Z.}~\bibnamefont {Hossain}},\ }\href@noop {} {\bibfield  {journal} {\bibinfo
   {journal} {Physical Review B}\ }\textbf {\bibinfo {volume} {103}},\ \bibinfo
  {pages} {L241112} (\bibinfo {year} {2021})}\BibitemShut {NoStop}%
\bibitem [{\citenamefont {Wang}\ \emph {et~al.}(2023)\citenamefont {Wang},
  \citenamefont {Li}, \citenamefont {Zhou}, \citenamefont {Chen}, \citenamefont
  {Wang}, \citenamefont {Yang}, \citenamefont {Zhou}, \citenamefont {Liao},
  \citenamefont {Weng},\ and\ \citenamefont {Wang}}]{wang2023structure}%
  \BibitemOpen
  \bibfield  {author} {\bibinfo {author} {\bibfnamefont {X.}~\bibnamefont
  {Wang}}, \bibinfo {author} {\bibfnamefont {B.}~\bibnamefont {Li}}, \bibinfo
  {author} {\bibfnamefont {L.}~\bibnamefont {Zhou}}, \bibinfo {author}
  {\bibfnamefont {L.}~\bibnamefont {Chen}}, \bibinfo {author} {\bibfnamefont
  {Y.}~\bibnamefont {Wang}}, \bibinfo {author} {\bibfnamefont {Y.}~\bibnamefont
  {Yang}}, \bibinfo {author} {\bibfnamefont {Y.}~\bibnamefont {Zhou}}, \bibinfo
  {author} {\bibfnamefont {K.}~\bibnamefont {Liao}}, \bibinfo {author}
  {\bibfnamefont {H.}~\bibnamefont {Weng}},\ and\ \bibinfo {author}
  {\bibfnamefont {G.}~\bibnamefont {Wang}},\ }\href@noop {} {\bibfield
  {journal} {\bibinfo  {journal} {arXiv preprint arXiv:2303.11894}\ } (\bibinfo
  {year} {2023})}\BibitemShut {NoStop}%
\bibitem [{\citenamefont {May}\ \emph {et~al.}(2023)\citenamefont {May},
  \citenamefont {Clements}, \citenamefont {Zhang}, \citenamefont {Hermann},
  \citenamefont {Yan},\ and\ \citenamefont {McGuire}}]{may2023coupling}%
  \BibitemOpen
  \bibfield  {author} {\bibinfo {author} {\bibfnamefont {A.~F.}\ \bibnamefont
  {May}}, \bibinfo {author} {\bibfnamefont {E.~M.}\ \bibnamefont {Clements}},
  \bibinfo {author} {\bibfnamefont {H.}~\bibnamefont {Zhang}}, \bibinfo
  {author} {\bibfnamefont {R.~P.}\ \bibnamefont {Hermann}}, \bibinfo {author}
  {\bibfnamefont {J.}~\bibnamefont {Yan}},\ and\ \bibinfo {author}
  {\bibfnamefont {M.~A.}\ \bibnamefont {McGuire}},\ }\href@noop {} {\bibfield
  {journal} {\bibinfo  {journal} {Physical Review Materials}\ }\textbf
  {\bibinfo {volume} {7}},\ \bibinfo {pages} {064406} (\bibinfo {year}
  {2023})}\BibitemShut {NoStop}%
\bibitem [{\citenamefont {Ge}\ \emph {et~al.}(2022)\citenamefont {Ge},
  \citenamefont {Jin},\ and\ \citenamefont {Zhu}}]{ge2022ferromagnetic}%
  \BibitemOpen
  \bibfield  {author} {\bibinfo {author} {\bibfnamefont {Y.}~\bibnamefont
  {Ge}}, \bibinfo {author} {\bibfnamefont {Y.}~\bibnamefont {Jin}},\ and\
  \bibinfo {author} {\bibfnamefont {Z.}~\bibnamefont {Zhu}},\ }\href@noop {}
  {\bibfield  {journal} {\bibinfo  {journal} {Materials Today Physics}\
  }\textbf {\bibinfo {volume} {22}},\ \bibinfo {pages} {100570} (\bibinfo
  {year} {2022})}\BibitemShut {NoStop}%
\bibitem [{\citenamefont {Barman}\ \emph {et~al.}(2020)\citenamefont {Barman},
  \citenamefont {Mondal}, \citenamefont {Pathak},\ and\ \citenamefont
  {Alam}}]{barman2020symmetry}%
  \BibitemOpen
  \bibfield  {author} {\bibinfo {author} {\bibfnamefont {C.~K.}\ \bibnamefont
  {Barman}}, \bibinfo {author} {\bibfnamefont {C.}~\bibnamefont {Mondal}},
  \bibinfo {author} {\bibfnamefont {B.}~\bibnamefont {Pathak}},\ and\ \bibinfo
  {author} {\bibfnamefont {A.}~\bibnamefont {Alam}},\ }\href@noop {} {\bibfield
   {journal} {\bibinfo  {journal} {Physical Review Materials}\ }\textbf
  {\bibinfo {volume} {4}},\ \bibinfo {pages} {084201} (\bibinfo {year}
  {2020})}\BibitemShut {NoStop}%
\bibitem [{\citenamefont {Turner}\ \emph {et~al.}(2012)\citenamefont {Turner},
  \citenamefont {Zhang}, \citenamefont {Mong},\ and\ \citenamefont
  {Vishwanath}}]{turner2012quantized}%
  \BibitemOpen
  \bibfield  {author} {\bibinfo {author} {\bibfnamefont {A.~M.}\ \bibnamefont
  {Turner}}, \bibinfo {author} {\bibfnamefont {Y.}~\bibnamefont {Zhang}},
  \bibinfo {author} {\bibfnamefont {R.~S.}\ \bibnamefont {Mong}},\ and\
  \bibinfo {author} {\bibfnamefont {A.}~\bibnamefont {Vishwanath}},\
  }\href@noop {} {\bibfield  {journal} {\bibinfo  {journal} {Physical Review
  B}\ }\textbf {\bibinfo {volume} {85}},\ \bibinfo {pages} {165120} (\bibinfo
  {year} {2012})}\BibitemShut {NoStop}%
\bibitem [{\citenamefont {Watanabe}\ \emph {et~al.}(2018)\citenamefont
  {Watanabe}, \citenamefont {Po},\ and\ \citenamefont
  {Vishwanath}}]{watanabe2018structure}%
  \BibitemOpen
  \bibfield  {author} {\bibinfo {author} {\bibfnamefont {H.}~\bibnamefont
  {Watanabe}}, \bibinfo {author} {\bibfnamefont {H.~C.}\ \bibnamefont {Po}},\
  and\ \bibinfo {author} {\bibfnamefont {A.}~\bibnamefont {Vishwanath}},\
  }\href@noop {} {\bibfield  {journal} {\bibinfo  {journal} {Science advances}\
  }\textbf {\bibinfo {volume} {4}},\ \bibinfo {pages} {eaat8685} (\bibinfo
  {year} {2018})}\BibitemShut {NoStop}%
\bibitem [{\citenamefont {Ono}\ and\ \citenamefont
  {Watanabe}(2018)}]{ono2018unified}%
  \BibitemOpen
  \bibfield  {author} {\bibinfo {author} {\bibfnamefont {S.}~\bibnamefont
  {Ono}}\ and\ \bibinfo {author} {\bibfnamefont {H.}~\bibnamefont {Watanabe}},\
  }\href@noop {} {\bibfield  {journal} {\bibinfo  {journal} {Physical Review
  B}\ }\textbf {\bibinfo {volume} {98}},\ \bibinfo {pages} {115150} (\bibinfo
  {year} {2018})}\BibitemShut {NoStop}%
\bibitem [{\citenamefont {Lv}\ \emph {et~al.}(2019)\citenamefont {Lv},
  \citenamefont {Qian},\ and\ \citenamefont {Ding}}]{lv2019angle}%
  \BibitemOpen
  \bibfield  {author} {\bibinfo {author} {\bibfnamefont {B.}~\bibnamefont
  {Lv}}, \bibinfo {author} {\bibfnamefont {T.}~\bibnamefont {Qian}},\ and\
  \bibinfo {author} {\bibfnamefont {H.}~\bibnamefont {Ding}},\ }\href@noop {}
  {\bibfield  {journal} {\bibinfo  {journal} {Nature Reviews Physics}\ }\textbf
  {\bibinfo {volume} {1}},\ \bibinfo {pages} {609} (\bibinfo {year}
  {2019})}\BibitemShut {NoStop}%
\bibitem [{\citenamefont {Ma}\ \emph {et~al.}(2020)\citenamefont {Ma},
  \citenamefont {Wang}, \citenamefont {Nie}, \citenamefont {Yi}, \citenamefont
  {Xu}, \citenamefont {Li}, \citenamefont {Jandke}, \citenamefont {Wulfhekel},
  \citenamefont {Huang}, \citenamefont {West} \emph
  {et~al.}}]{ma2020emergence}%
  \BibitemOpen
  \bibfield  {author} {\bibinfo {author} {\bibfnamefont {J.}~\bibnamefont
  {Ma}}, \bibinfo {author} {\bibfnamefont {H.}~\bibnamefont {Wang}}, \bibinfo
  {author} {\bibfnamefont {S.}~\bibnamefont {Nie}}, \bibinfo {author}
  {\bibfnamefont {C.}~\bibnamefont {Yi}}, \bibinfo {author} {\bibfnamefont
  {Y.}~\bibnamefont {Xu}}, \bibinfo {author} {\bibfnamefont {H.}~\bibnamefont
  {Li}}, \bibinfo {author} {\bibfnamefont {J.}~\bibnamefont {Jandke}}, \bibinfo
  {author} {\bibfnamefont {W.}~\bibnamefont {Wulfhekel}}, \bibinfo {author}
  {\bibfnamefont {Y.}~\bibnamefont {Huang}}, \bibinfo {author} {\bibfnamefont
  {D.}~\bibnamefont {West}}, \emph {et~al.},\ }\href@noop {} {\bibfield
  {journal} {\bibinfo  {journal} {Advanced Materials}\ }\textbf {\bibinfo
  {volume} {32}},\ \bibinfo {pages} {1907565} (\bibinfo {year}
  {2020})}\BibitemShut {NoStop}%
\end{thebibliography}%

\end{document}